\documentclass[a4paper,twocolumn,superscriptaddress,prb,showpacs]{revtex4}
\usepackage{graphicx}
\usepackage{epstopdf}
\usepackage{amssymb}

\begin{document}

\title{Decoherence, Autler-Townes effect, and dark states in two-tone driving of a three-level superconducting system}



\author{Jian Li}
\email{jianli@ltl.tkk.fi}

\author{G. S. Paraoanu}
\affiliation{Low Temperature Laboratory, Aalto University, PO Box 15100, FI-00076 AALTO,
Finland}

\author{Katarina Cicak}
\affiliation{National Institute of Standards and Technology, 325 Broadway, Boulder,
Colorado 80305, USA}

\author{Fabio Altomare}
\altaffiliation{Present affiliation: D-Wave Systems Inc., 100-4401 Still Creek Drive,
Burnaby, British Columbia V5C 6G9, Canada}

\author{Jae I. Park}
\affiliation{National Institute of Standards and Technology, 325 Broadway, Boulder,
Colorado 80305, USA}

\author{Raymond W. Simmonds}
\affiliation{National Institute of Standards and Technology, 325 Broadway, Boulder,
Colorado 80305, USA}

\author{Mika A. Sillanp\"a\"a}
\affiliation{Low Temperature Laboratory, Aalto University, PO Box 15100, FI-00076 AALTO,
Finland}

\author{Pertti J. Hakonen}
\affiliation{Low Temperature Laboratory, Aalto University, PO Box 15100, FI-00076 AALTO,
Finland}


\begin{abstract}

We present a detailed theoretical analysis of a multi-level quantum system coupled to two radiation fields and subject to decoherence. We concentrate on an effect known from quantum optics as Autler-Townes splitting, which has been recently demonstrated experimentally [M. A. Sillanp\"a\"a {\em et al.}, Phys. Rev. Lett. {\bf 103}, 193601 (2009)] in a superconducting phase qubit. In the three-level approximation, we derive analytical solutions and describe how they can be used to extract  the decoherence rates and to account for the measurement data. Better agreement with the experiment can be obtained by extending this model to five levels. Finally, we investigate the stationary states created in  the experiment and show that their structure is close to that of dark states.
\end{abstract}


\maketitle


\section{Introduction}
\label{intro}

During the last decade, the field of superconducting qubits has advanced tremendously \cite{review}. These systems behave quantum-mechanically, and can be regarded as tunable artificial atoms. Different from their natural counterparts, they couple strongly with  the environment and as a result they have shorter coherence times. On the other hand, stronger coupling has its own advantages, such as easier addressability and shorter gate times.

A number of quantum phenomena become manifest when atoms interact with electromagnetic radiation. Among these, electromagnetically induced transparency in atomic media allows spectacular effects such as the reduction of the group velocity  of light to a few meters per second \cite{hau}  or even a complete stop. This phenomenon requires two fields (the "probe" and the "coupling" - or "pump"- field) and a three-level atom. The two fields are typically on-resonance (or not far off-resonance) from the two transitions. This effect can be used for coherent storage of optical information \cite{stop}, for the realization of quantum repeaters \cite{repeaters}, for enhanced photon-photon interactions \cite{single}, and for setting up table-top cosmological experiments such as the creation of event horizons \cite{ulf}. These phenomena are not restricted to atomic physics, but solid-state systems can also be used: for example, ultraslow light propagation  has been already demonstrated in crystals \cite{crystal}. Electromagnetically induced transparency has its roots in the destructive interference of probability amplitudes of the state that is coupled to both fields: as a result, the "probe" field is no longer absorbed and the atom becomes "trapped" in a "dark" state (a superposition formed by the other two states), a phenomenon called "coherent population trapping". This phenomenon could find applications as a way to precisely prepare initial states in future quantum computers. In solid-state systems, this effect has so far been demonstrated with donor-bound spins in GaAs \cite{gaas}, with nitrogen-vacancy centers in diamond \cite{nv}, and with single spins in quantum dots \cite{xu}.

Superconducting quantum circuits can be operated as well as three-level systems interacting with two radiation fields on-resonant with the two transitions. So far, this has been realized with phase qubits  \cite{mika,bbn}, transmons \cite{wallraff}, and flux qubits \cite{tsai}. From the spectra obtained in these experiments it is sometimes difficult to distinguish between electromagnetically induced transparency (whose origin is quantum interference) and the Autler-Townes effect \cite{AT} (which is due to the shift of the resonance of the transition frequency which is probed). As a rule of thumb to separate the two effects\cite{sanders}, if the coupling-field Rabi frequency (denoted by $\Omega_c$ in this paper) is much smaller than the spectral linewidth of the darkened transition and in absorption spectroscopy we see a sharp dip formed within the linewidth, then we have electromagnetically induced transparency. If the Rabi frequency of the coupling  field is much larger then the linewidth, we have the Autler-Townes effect, which appears as the splitting of the spectral line into two lines of width similar to the initial one and separated by $\approx \Omega_{c}$. Evidence for coherent population trapping and electromagnetically induced transparency is so far only indirect: there is yet no quantum tomography experiment performed to truly measure the superposition mentioned above (see the previous paragraph). In the experiment of Ref. [\onlinecite{mika}] analyzed in this paper the power of the coupling field is such that $\Omega_c$ is larger than the linewidth, making the second effect dominant. We will therefore refer  to the phenomenon described below as the Autler-Townes effect in a superconducting qubit.

We consider here a phase qubit: a system consisting of a superconducting loop interrupted by  a Josephson junction \cite{martinis}. The three lowest energy levels are denoted by $|0\rangle$, $|1\rangle$, and $|2\rangle$, and the corresponding transitions can be driven\cite{alsomartinis} by microwave radiation fields with frequencies of  the order of a few gigahertz.  In atomic physics \cite{EITatoms}, the states typically form a $\Lambda$ configuration, with $|0\rangle$ and $|2\rangle$ being metastable hyperfine or Zeeman levels, while the state $|1\rangle$ is usually an excited electronic state that decays at a faster rate. The situation in the case of phase qubits differs: since the direct transition $|0\rangle \rightarrow |2\rangle$ is suppressed in these systems due to relatively low anharmonicity, the most straightforward operation of the phase qubits is as ladder systems, that is, driving only on the allowed single-photon transitions $|0\rangle \rightarrow |1\rangle$ and $|1\rangle \rightarrow |2\rangle$.  It is also possible to operate these systems in the $\Lambda$ configuration \cite{bbn}, by driving two-photon virtual transitions from $|0\rangle$ to $|2\rangle$ with an intense microwave field at the frequency $f_{02}/2$. This frequency can be sufficiently detuned from the transition $f_{01}$ that the direct excitation into the state $|1\rangle$ is small. Depending on the sample design, one can measure either the level population of  the qubit  (if it can be addressed directly), or the scattered radiation.  For example, in Ref. [\onlinecite{tsai}] the qubit is embedded in a one-dimensional transmission line, and what has been measured is the absorption of a probe microwave signal in resonance with the $|0\rangle\rightarrow|1\rangle$ transition. What is seen  is that this quantity becomes large (close to 1 within a few percentage points) when a more intense field is applied to the transition $|1\rangle\rightarrow|2\rangle$.

Future quantum-information processing devices based on superconducting circuit architectures could make use of microwave-controlled states in three-level systems. In addition to the applications listed above, specific to the field of circuit quantum electrodynamics one could employ three-level effects for the coherent and tunable excitation of microwaves in coplanar waveguides \cite{1demission}, as an alternative way to measure the qubit's decoherence rates \cite{decoherence}, for cooling \cite{cooling} and single-photon generation \cite{sp}, as a single-atom media quantum amplifier \cite{astafiev}, and as quantum switches \cite{sw}.

This work is organized as follows: in Section \ref{multi_level} we introduce the phase qubit and derive its corresponding Hamiltonian. In Section \ref{master_eq} we write the general form of the Lindblad master equation for the phase qubit approximated as a three-level system, including relaxation and dephasing in our model. A full derivation of  the relaxation term and of the dephasing term can be found in Appendices \ref{appendixa} and \ref{appendixb}, respectively. The three-level model can be easily extended to account for spurious excitations on the next two levels, and the resulting five-level model is briefly presented in Section \ref{5levels}. In Section \ref{dephasing} we then explain how to extract the dephasing rates corresponding to the three levels from the occupation probabilities under continuous irradiation (the relaxation rates are obtained from independent pulsed measurements). Next, we present a number of approximate analytical results (Section \ref{analytical}) for the Autler-Townes splitting and for the occupation probabilities. In Section \ref{comparison} we present the experimental data for the Autler-Townes effect and we compare them with the theoretical predictions based on the parameters determined in Section \ref{dephasing}. We show that the approximate analytical results presented in Section \ref{analytical} provide a good fit to the data. Better agreement can be obtained with numerical simulations for the full five-level model of Section \ref{5levels}.  Finally, in Section \ref{dark} we calculate the fidelities of the steady states relative to ideal dark states. Details about the structure of the steady states are delegated to Appendix \ref{appendixc}.


\section{A three-level artificial atom}
\label{multi_level}

\begin{figure}[h]
\includegraphics[width=6cm]{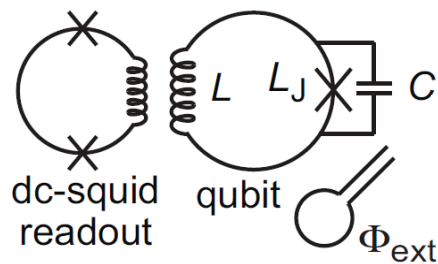}
\caption{Schematic of the phase qubit, consisting of a single junction with capacitance $C$ and Josephson inductance $L_{J}$ inserted into a superconducting loop of inductance $L$. A nearby dc-SQUID is used for single-shot readout and an on-chip external line is employed to bias the qubit with external flux $\Phi_{\rm ext}$ with both dc and rf components, $\Phi_{\rm ext} = \Phi_{\rm dc} + \Phi_{\rm rf}(t)$.} \label{fig_SQUID}
\end{figure}

In this section we introduce our phase qubit \cite{martinis}, and analyze it theoretically in the three-level approximation.  The phase qubit can be pictured as an rf-SQUID (see Fig. \ref{fig_SQUID}), consisting of a single Josephson junction with critical current $I_c$ and capacitance $C$, which has been inserted into a superconducting loop with inductance $L$. We denote by $\Phi_0$ the flux quantum, $ \Phi_{0} = h/2e = 2.067 \times 10^{-15}$ Wb. The Josephson energy of the junction is then $E_{J} = (\Phi_{0}/2\pi) I_{c}$; the application of a phase difference $\varphi$ requires an energy $E_{J}(1-\cos\varphi )$, and  can be regarded as resulting from a nonlinear inductance $L_J (\varphi) =  L_{J}/\cos \varphi$, where $L_J  = (\Phi_0 / 2\pi)^2 / E_J = \Phi_0 / (2\pi I_{c})$ can be understood as the Josephson inductance in the limit of small phase differences.

The classical Lagrangian of the rf-SQUID can be written in the standard form as kinetic energy minus potential energy,
\begin{equation}
{\cal L}  = \frac{1}{2}C\dot{\Phi}^2 - \left[\frac{(\Phi - \Phi_{\rm ext})^2}{2L} - E_J \cos\left(2\pi\frac{\Phi}{\Phi_0}\right) \right], \label{eq_lagrangian}
\end{equation}
where $\Phi/\Phi_{0}$ is the superconducting phase difference across the junction and  $\Phi_{\rm ext} = \Phi_{\rm dc} + \Phi_{\rm rf}(t)$ is an external magnetic flux with both dc and rf components applied though an on-chip flux bias coil. By using the Legendre transformation, we obtain the Hamiltonian
\begin{eqnarray}
H &=& \frac{\partial {\cal L}}{\partial \dot{\Phi}} \dot{\Phi} - {\cal L} \nonumber \\
&=& \frac{Q^2}{2C} + \frac{(\Phi - \Phi_{\rm ext})^2}{2L} - E_J\cos\left( 2\pi\frac{\Phi}{\Phi_0} \right) \label{eq_hamiltonian1} \\
&\approx& \frac{Q^2}{2C} + \frac{(\Phi - \Phi_{\rm dc})^2}{2L} - E_J\cos\left( 2\pi\frac{\Phi}{\Phi_0} \right) - \frac{\Phi\Phi_{\rm rf}(t)}{L} , \nonumber
\end{eqnarray}
where the canonically conjugate variable of the flux $\Phi$ is the charge $Q = \partial{\cal L} / \partial\dot{\Phi}=C\dot{\Phi}$ accumulated on the capacitor $C$, and we assume $|\Phi_{dc}|\gg|\Phi_{rf}(t)|$. Let us first study the time-independent part $H'$ of $H$, defined by $H = H' -  \Phi\Phi_{\rm rf}(t) /L$; we will return to the time-dependent part toward the end of this section. $H'$ can be conveniently written as $H' = Q^2 /2C + E_{\rm pot}$, where the potential energy is
\begin{equation}
E_{\rm pot} =  \frac{E_L}{2}(\phi - \phi_{\rm dc})^2 - E_J\cos\phi , \label{eq_potential1}
\end{equation}
with the definitions $E_L=(\Phi_0/2\pi)^2/2L$, $\phi = 2\pi\Phi /\Phi_0$ and $\phi_{\rm dc}= 2\pi\Phi_{\rm dc}/\Phi_0$. Also, the loop inductance $L>L_J$ is chosen so that local minima are formed in the potential.

In order to analyze the quantum states trapped in one local minimum of the potential given by Eq. (\ref{eq_potential1}), we expand Eq. (\ref{eq_potential1}) as a Taylor series around the first local minimum $\phi_m$ (implying that $0 \lesssim \phi_m \lesssim 2\pi$):
\begin{eqnarray}
E_{\rm pot} &\approx& E_{\rm pot}(\phi_m) + E'_{\rm pot}(\phi_m)\Delta\phi + \frac{1}{2}E''_{\rm pot}(\phi_m)\Delta\phi^2 \nonumber \\
&& + \frac{1}{6}E'''_{\rm pot}(\phi_m)\Delta\phi^3 + \cdots . \label{eq_taylor}
\end{eqnarray}
Here $\Delta\phi = \phi - \phi_m$, and the first three derivatives are
\begin{eqnarray}
&& E'_{\rm pot}(\phi_m) = E_L(\phi_m - \phi_{\rm dc}) + E_J\sin\phi_m , \nonumber \\
&& E''_{\rm pot}(\phi_m) = E_L + E_J\cos\phi_m , \nonumber \\
&& E'''_{\rm pot}(\phi_m) = -E_J\sin\phi_m .\nonumber
 \label{eq_derivatives1}
\end{eqnarray}

To obtain the local minimum $\phi_m$, we need to solve the equation $E'_{pot}(\phi_m) = 0$.  This is not solvable analytically, but the  quadratic expansion of the sine function around $3\pi/2$,
\begin{equation}
\sin\phi_m \approx -1 + \frac{1}{2}\left( \phi_m - \frac{3\pi}{2} \right)^2 ,
\end{equation}
gives a good-enough approximation of the first positive-value local minimum $\phi_m$,
\begin{eqnarray}
\phi_m &=& \frac{3\pi}{2} - \frac{E_L}{E_J} + \sqrt{2 + \frac{E_L^2}{E_J^2} + \frac{E_L}{E_J}(2\phi_{\rm dc} - 3\pi)} \nonumber \\
&=& \frac{3\pi}{2} - \beta^{-1} + \sqrt{2 + (2\phi_{\rm dc} - 3\pi)\beta^{-1} + \beta^{-2}} , \label{eq_localmin}
\end{eqnarray}
where $\beta = (2 \pi /\Phi_{0}) (L/I_{c}) = L/L_J = E_J / E_L$. Since $\phi_m$ should be a real number, $2 + (2\phi_{\rm dc} - 3\pi)\beta^{-1} + \beta^{-2} > 0$ must be satisfied, thus
\begin{equation}
\phi_{\rm dc} > \frac{3\pi}{2}-\frac{1}{2\beta}-\beta . \label{eq_phidc}
\end{equation}

\begin{figure}[h]
\includegraphics[width=7cm]{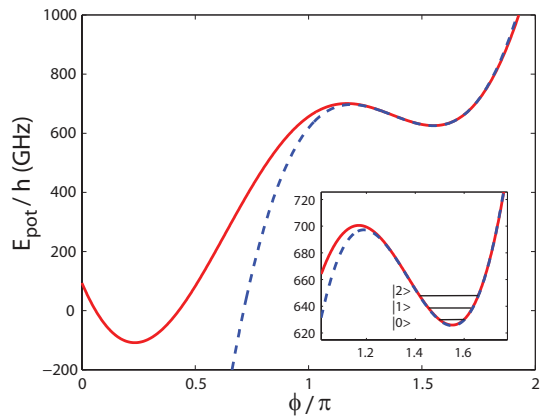}
\caption{(color online). Potential energy of the rf-SQUID for typical parameters $L_J = 276$ pH, $L = 690$ pH, and $\phi_{\rm dc} = 0.766\pi$. The minimum is located at $\phi_m \approx 1.551\pi$. The (red) solid curve is the exact potential energy calculated with Eq. (\ref{eq_potential1}); the (blue) dashed curve is the cubic approximation given by Eq. (\ref{eq_potential2}). The inset shows the energies of the first three metastable states $|0\rangle, |1\rangle$, and $|2\rangle$  formed in the well centered around the local minimum $\phi_{m}$.} \label{fig_potential}
\end{figure}

Having found the local minimum $\phi_m$, we can then rewrite Eq. (\ref{eq_taylor}) explicitly, using also $\cos (\varphi_m) = \cos (\varphi_m - 3\pi /2 + 3\pi /2) \approx \cos (\varphi_m - 3\pi /2 )\cos (3\pi/2) - \sin (3\pi /2)\sin (\varphi_m - 3\pi /2 ) \approx \varphi_{m} - 3\pi /2$,
\begin{eqnarray}
E_{\rm pot}(\Delta\phi) &\approx& \frac{E_L}{2} (\phi_m - \phi_{\rm dc})^2 - E_J (\phi_m - 3\pi/2) \nonumber \\
&& + \frac{E_J}{2}\sqrt{2 + (2\phi_{\rm dc} - 3\pi)\beta^{-1} + \beta^{-2}} \Delta\phi^2 \nonumber \\
&& + \frac{E_L}{12} [ 2\sqrt{2 + (2\phi_{\rm dc} - 3\pi)\beta^{-1} + \beta^{-2}} \nonumber \\
&& \ \ \ \ \ \ \ \ - (2\phi_{\rm dc} - 3\pi + 2\beta^{-1}) ] \Delta\phi^3 . \label{eq_potential2}
\end{eqnarray}
The comparison between this cubic potential and the original potential is shown in Fig. \ref{fig_potential}. We see that around the local minimum, the cubic potential approximation is excellent, which allows us to use it for extracting the energy levels.

The last step is to derive the quantum Hamiltonian in the three-level  approximation. To simplify the notations, we can define
\begin{eqnarray}
&& \lambda = \left[ 2 + (2\phi_{\rm dc} - 3\pi)\beta^{-1} + \beta^{-2} \right]^{1/2} , \nonumber \\
&& \xi = 3\pi - 2 (\phi_{\rm dc} + \beta^{-1}) , \nonumber \\
\end{eqnarray}
and we rewrite Eq. (\ref{eq_potential2}) as (dropping terms independent of $\Delta\phi$)
\begin{equation}
E_{\rm pot} \approx \frac{E_{J}\lambda}{2}\Delta\phi^2 + \frac{E_{L}(2\lambda + \xi )}{12}\Delta\phi^3 . \label{eq_potential03}
\end{equation}
As one can see from the expression above, the energies that define the shape of the metastable well in this approximation are $E_{J}\lambda$ and $E_{L}(2\lambda + \xi )$. Also, since the value $\phi_m$ will play no role from now on, we will make the convention that the variables $\phi$ and $\Phi =  \frac{\Phi_0}{2\pi}\phi$ are measured from $\phi_m$ as a reference point. With this convention, we can rewrite Eq. (\ref{eq_potential03}) in terms of the fluxes $\Phi$,
\begin{equation}
E_{\rm pot} \approx \frac{1}{2L_J^\ast}\Phi^2 +  \frac{1}{2\Phi_0L^\ast}\Phi^3 , \label{eq_potential3}
\end{equation}
where $L_J^\ast = L_J / \lambda$ and $L^\ast = 3L / \pi(2\lambda + \xi)$.

Summing up the results, the total time-independent Hamiltonian $H'$ can be written as
\begin{equation}
H' = \frac{Q^2}{2C} + \frac{\Phi^2}{2L_J^\ast} + \frac{\Phi^3}{2\Phi_0 L^\ast} = H_0 + \frac{\Phi^3}{2\Phi_0 L^\ast}, \label{eq_hamiltonian2}
\end{equation}
where
\begin{equation}
H_{0} = \frac{Q^2}{2C} + \frac{\Phi^2}{2L_J^\ast} \label{eq_hamiltonian02}
\end{equation}
describes a harmonic oscillator.

We can now go further and quantize our Hamiltonian. We immediately notice that  the canonically conjugate variable of $Q$ is $\Phi$. Thus the commutation relation $[\Phi, Q] = i\hbar$ can be postulated. We can now introduce the creation and annihilation operators for the harmonic oscillator Eq. (\ref{eq_hamiltonian02}), $a^\dag$ and $a$, defined by
\begin{equation}
\Phi  =  \sqrt{\frac{\hbar}{2C\omega_{o}}}(a^\dag +a),\ \ \
 Q = \sqrt{\frac{\hbar C\omega_{0}}{2}}(a^\dag -a),
 \end{equation}
 and satisfying the algebra $[a,a] = [a^\dag, a^\dag] = 0$, $[a, a^\dag] = 1$ (here $\omega_0 = \sqrt{1 / L_J^\ast C}$ is the Josephson plasma frequency). Thus
\begin{equation}
H_0 =\hbar\omega_0 \left( a^\dag a + 1/2 \right),
\end{equation}
with eigenvectors $|n\rangle_0$,
$H_{0}|n\rangle_0 = \hbar\omega_0 (n + 1/2)|n\rangle_0$.

To proceed with the full time-independent Hamiltonian $H'$, we note that around the local minimum, $\Phi / \Phi_0 \ll 1$. Therefore, we can treat the second term on  the right-hand side of Eq. (\ref{eq_potential3}) and Eq. (\ref{eq_hamiltonian2}) as a perturbation.

Truncating to the lowest three unperturbed eigenstates $|n\rangle_0$ of $H_0$ we get Eq. (\ref{eq_hamiltonian2}) in the matrix form
\begin{equation}
H' = \left[ \begin{array}{ccc}
        \hbar\omega_0/2 & 3\eta & 0 \\
        3\eta & 3\hbar\omega_0/2 & 6\sqrt{2}\eta \\
        0 & 6\sqrt{2}\eta & 5\hbar\omega_0/2
\end{array} \right] , \label{eq_hamiltonian3}
\end{equation}
where we defined $\eta = (\hbar/2C\omega_0)^{3/2} / 2 \Phi_{0}L^\ast$.  The perturbation $\frac{\Phi^3}{2\Phi_0 L^\ast}$ thus couples only the states $|0\rangle_0$ and  $|1\rangle_0$ with strength $3\eta$, and $|1\rangle_0$ and $|2\rangle_0$, with strength $6\sqrt{2}\eta$. The matrix Eq. (\ref{eq_hamiltonian3}) can be diagonalized exactly. The resulting eigenvalues are $\hbar\omega_{0}/2 - 11\eta^{2}/\hbar\omega_0$, $3\hbar\omega_{0}/2 - 71\eta^{2}/\hbar\omega_{0}$, and $5\hbar\omega_{0}/2 - 191\eta^{2}/\hbar\omega_0$. For the fabrication parameters of our sample (see {\em e.g.} Fig. \ref{fig_potential}), $\eta$ is only about $2\%$ of $\hbar\omega_0$; hence the eigenstates $|n\rangle$ of $H'$ are quite close to the unperturbed eigenstates $|n\rangle_0$.

We now turn to the full Hamiltonian Eq. (\ref{eq_hamiltonian1}), including the time-dependent term $\Phi\Phi_{\rm rf}/L$, and to avoid carrying over the zero-point energy in the equations, we make the convention that all the energies are measured from the ground-state level. We first give the numerical form of the Hamiltonian suitable for the analysis of our experiments, which is obtained by calculating the matrix elements of the time-dependent terms in the  basis $|n\rangle$ with $\eta \approx 0.02\hbar\omega_0$; we obtain
\begin{eqnarray}
H(t) &=& H' - \frac{\Phi\Phi_{\rm rf}(t)}{L} \nonumber \\
&\approx& \hbar \left[ \begin{array}{ccc}
       0 & 0.69 g(t) & 0.02 g(t) \\
       0.69 g(t) & \omega_{10} & g(t) \\
       0.02 g(t) & g(t) & \omega_{10} + \omega_{21}
\end{array} \right] , \label{eq_hamiltonian5}
\end{eqnarray}
where
\begin{equation}
g(t) = -\frac{\Phi_{\rm rf}(t)}{L}\sqrt{\frac{1}{C \hbar\omega_0 }}
\end{equation}
is a coupling constant (measured in hertz), and the transition frequencies are
\begin{eqnarray}
\omega_{10} = 0.976\omega_0 , \ \ \omega_{21} = 0.952\omega_0 . \label{omegas}
\end{eqnarray}
The frequencies $\omega_{10}$ and $\omega_{21}$ can be found spectroscopically. For our sample we find $\omega_{10}=2\pi \times 8.135$ GHz and $\omega_{21}=2\pi \times 7.975$ GHz. The experimentally determined relative anharmonicity $(\omega_{10} -\omega_{21})/(\omega_{10}+\omega_{21})$ is $9.9\times 10^{-3}$, in reasonable agreement with the value $12\times 10^{-3}$ obtained by using Eqs. (\ref{omegas}).

As a comparison, if we completely neglect the perturbation ($\eta\rightarrow 0$),  Eq. (\ref{eq_hamiltonian5}) simplifies to
\begin{equation}
H_{0} - \frac{\Phi\Phi_{rf}(t)}{L} = \hbar \left[ \begin{array}{ccc}
       0 & \frac{1}{\sqrt{2}} g(t) & 0 \\
        \frac{1}{\sqrt{2}} g(t) & \omega_0 & g(t) \\
       0 & g(t) & 2\omega_0
\end{array} \right] . \label{eq_hamiltonian50}
\end{equation}
This expression clearly reveals the structure of the circuit Hamiltonian as a ladder system, with interlevel coupling strengths given by $g(t)/\sqrt{2}$ for the $|0\rangle \rightarrow |1\rangle$ transition and $g(t)$ for the $|1\rangle \rightarrow |2\rangle$ transition. Also from Eq. (\ref{eq_hamiltonian50}) as well as from  Eq. (\ref{eq_hamiltonian5}) one sees that if the qubit is weakly anharmonic, the transition $|0\rangle \rightarrow |2\rangle$ cannot be driven directly. However, at high enough powers, two-photon transitions between these two levels are observed experimentally.

In our Autler-Townes splitting experiments, $\Phi_{rf}(t)$ consists of two microwave tones: the weak probe tone $\Phi_p\cos(\omega_p t)$ and the strong coupling tone $\Phi_c\cos(\omega_c t)$. In Eq. (\ref{eq_hamiltonian5}) we can write $g(t)$ as $g(t) = g_c\cos(\omega_c t) + g_p\cos(\omega_p t)$, where the coupling and probe amplitudes read
\begin{eqnarray}
g_c = -\frac{\Phi_c}{L}\sqrt{\frac{1}{C\hbar\omega_0}} , \ \ g_p = -\frac{\Phi_p}{L}\sqrt{\frac{1}{C\hbar\omega_0}} . \nonumber
\end{eqnarray}

The Hamiltonian Eq. (\ref{eq_hamiltonian5}) clearly shows that both microwave tones are coupled to $|0\rangle\leftrightarrow |1\rangle$, $|1\rangle\leftrightarrow |2\rangle$ and $|0\rangle\leftrightarrow |2\rangle$ transitions (cross couplings). The cross couplings introduce additional complications to any analytical study of this system. In the next section we will show that they can be eliminated by changing to suitable rotating reference frames. This allows us to then derive relatively simple analytical results.


\section{The master equation}
\label{master_eq}

The driven three-level qubit is inevitably coupled to the external electromagnetic environment, which causes relaxation and dephasing of the system.  One can easily extend the well known spin-boson model \cite{Carmichael, Breuer} of dissipation in two-level systems to this three-level case. See also Ref. [\onlinecite{manylevels}] for a similar treatment of decoherence in a multilevel qubit.

The standard Markovian master equation for the Schr\"odinger-picture density matrix $\rho_{\rm S}$ of the phase qubit has the form
\begin{equation}
\dot{\rho_{\rm S}} = -\frac{i}{\hbar}\left[ H, \rho_{\rm S} \right] + {\cal L}[\rho_{\rm S}] , \label{eq_master_sp}
\end{equation}
where $H$ is given by Eq. (\ref{eq_hamiltonian5}), and ${\cal L}$ is the total Liouville superoperator.
There are two contributions to ${\cal L}$: the first is due to relaxation, and we will call it ${\cal L}_{\rm rel}$, and the second is caused by fluctuations of the energy levels, which we will refer to
as pure dephasing, ${\cal L}_{\rm dep}$. Thus ${\cal L}={\cal L}_{\rm rel} + {\cal L}_{\rm dep}$.
The reason for calling  ${\cal L}_{\rm dep}$ "pure" is that, as we will see below, relaxation produces interlevel dephasing as well. A microscopic derivation of these Liouvilleans is given in
Appendix \ref{appendixa} and \ref{appendixb}. One important difference between relaxation and pure dephasing is that the first results from energy exchange with the environment, while the second is caused by energy-conserving virtual processes. This allows us to determine the parameters entering ${\cal L}_{\rm rel}$ in independent experiments, by exciting the system and measuring how long it takes to decay into the environment (more details in Section \ref{dephasing}).

The action of the Liouville superoperator for relaxation ${\cal L}_{\rm rel}$ is given by
\begin{eqnarray}
{\cal L}_{\rm rel}[\rho_{\rm S}] &=& \frac{\Gamma_{10}}{2}\left( 2\sigma_{01}\rho_{\rm S}\sigma_{10} - \sigma_{11}\rho_{\rm S} - \rho_{\rm S}\sigma_{11} \right) \nonumber \\
&& + \frac{\Gamma_{21}}{2}\left( 2\sigma_{12}\rho_{\rm S}\sigma_{21} - \sigma_{22}\rho_{\rm S} - \rho_{\rm S}\sigma_{22} \right) \nonumber \\
&& + \kappa \left( \sigma_{01}\rho_{\rm S}\sigma_{21} + \sigma_{12}\rho_{\rm S}\sigma_{10} \right) ,
\label{eq_relax}
\end{eqnarray}
where $\kappa = \sqrt{\Gamma_{10}\Gamma_{21}}$. Here, $\sigma_{ij} = |i\rangle \langle j|$ and  the interlevel relaxation rates between $|1\rangle \rightarrow |0\rangle$ and $|2\rangle \rightarrow |1\rangle$ are denoted as $\Gamma_{10}$ and $\Gamma_{21}$, respectively. Note also that for a three-level system, ${\cal L}_{\rm rel}$ is not just a simple generalization of the well-known expression for two-level systems, and a mixing term with effective decay constant $\kappa$ appears -- see the last term in Eq. (\ref{eq_relax}). This term however can be neglected after performing a rotating wave approximation (see below). A complete derivation of the relaxation Liouvillean is given in Appendix \ref{appendixa}.

For the pure dephasing part of the Liouvillean we will use the general expression
\begin{equation}
{\cal L}_{\rm dep} [\rho_{\rm S}] =
- \sum_{j\neq k} \frac{\gamma^{\varphi}_{jk}}{2} \sigma_{jj}\rho_{\rm S}\sigma_{kk}, \label{eq_dephs}
\end{equation}
where  $\gamma^{\varphi}_{jk}=\gamma^{\varphi}_{kj}$ are pure interlevel dephasing rates, and $j,k \in \{0,1,2\}$. A derivation of this expression can be found in Appendix \ref{appendixb}.



It is convenient to bring the system into a doubly-rotating reference frame defined by the operator
\begin{equation}
U =  \left[ \begin{array}{ccc}
       1 & 0 & 0 \\
       0 & e^{-i\omega_p t} & 0 \\
       0 & 0 & e^{-i(\omega_p+\omega_c)t}
\end{array} \right] . \label{frame}
\end{equation}
Under this unitary transformation the master equation preserves its structure, $\dot{\rho} = -(i/\hbar) [\tilde{H},\rho] + \tilde{\cal L}[\rho]$, where $\rho =  U^\dag \rho_{\rm S} U$,  the Hamiltonian transforms as $\tilde{H} = U^\dag HU + i\hbar (\partial U^\dag/\partial t)U$, and the Liouvillean as $\tilde{\cal L}[\rho] = U^{\dag} {\cal L}[\rho_{\rm S}] U$.
Performing a rotating wave approximation (RWA) to drop terms oscillating with frequencies $\omega_p$, $\omega_c$ and $\omega_p+\omega_c$, we obtain
\begin{widetext}
\begin{equation}
H^{\rm (RWA)}/\hbar = \left[ \begin{array}{ccc}
       0 & 0.69\left( g_p + g_c e^{-i\delta t} \right)/2 & 0 \\
       0.69\left( g_p + g_c e^{i\delta t} \right)/2 & \Delta_p & \left( g_p e^{i\delta t} + g_c \right)/2 \\
       0 & \left( g_p e^{-i\delta t} + g_c \right)/2 & \Delta_p + \Delta_c
\end{array} \right] , \label{eq_hamiltonian_rf}
\end{equation}
\end{widetext}
where the detunings $\Delta_p = \omega_{10} - \omega_p$, $\Delta_c = \omega_{21} - \omega_c$, and $\delta = \omega_p - \omega_c$. The oscillating terms in Eq. (\ref{eq_hamiltonian_rf}) average out to zero on the time scale of our experiment, thus for the calculation of steady state ($\dot{\rho} = 0$) we can neglect them \cite{Puri}, thereby obtaining the final matrix representation for the effective Hamiltonian:
\begin{equation}
H^{\rm (eff)}/\hbar = \left[ \begin{array}{ccc}
0 & \Omega_p/2 & 0 \\
\Omega_p/2 & \Delta_p & \Omega_c/2 \\
0 & \Omega_c/2 & \Delta_p + \Delta_c
\end{array} \right] , \label{effective_hamiltonian}
\end{equation}
where the probe and the coupling Rabi frequencies are defined as $\Omega_p = 0.69 g_p$ and $\Omega_c = g_c$, respectively.

The relaxation Liouvillean in the doubly-rotating frame reads

\begin{eqnarray}
\tilde{{\cal L}}_{\rm rel}[\rho] &=& \frac{\Gamma_{10}}{2}\left( 2\sigma_{01}\rho\sigma_{10} - \sigma_{11}\rho - \rho\sigma_{11} \right) \nonumber \\
&& + \frac{\Gamma_{21}}{2}\left( 2\sigma_{12}\rho\sigma_{21} - \sigma_{22}\rho - \rho\sigma_{22} \right) \nonumber \\
&& + \kappa \left( e^{-i\delta t}\sigma_{01}\rho\sigma_{21} + e^{i\delta t}\sigma_{12}\rho\sigma_{10} \right), \nonumber \\
 \label{eq_liouville}
\end{eqnarray}
and again we will neglect the oscillating term. Also, it is easy to check that the form Eq. (\ref{eq_dephs}) of the pure dephasing part is unchanged in the doubly-rotating frame. Thus decoherence is described by the following effective Liouvillean,
\begin{eqnarray}
{\cal L}^{({\rm eff})} [\rho] &=& \sum_{j\in \{1,2\}}
\frac{\Gamma_{j,j-1}}{2}
(2\sigma_{j-1,j}\rho\sigma_{j,j-1} - \sigma_{jj}\rho -
\rho\sigma_{jj})
\nonumber \\
&& - \sum_{j,k\in\{0,1,2\};j\neq k}\frac{\gamma^{\varphi}_{jk}}{2} \sigma_{jj}\rho \sigma_{kk},\label{eq_liouville_rf}
\end{eqnarray}
or in explicit matrix form
\begin{widetext}
\begin{equation}
{\cal L}^{\rm (eff)}[\rho] = \frac{1}{2} \left[ \begin{array}{ccc}
        2\Gamma_{10}\rho_{11} & -(\Gamma_{10} + \gamma^{\varphi}_{10})\rho_{01} & -(\Gamma_{21}+ \gamma^{\varphi}_{20})\rho_{02} \\
        -(\Gamma_{10} + \gamma^{\varphi}_{10})\rho_{10} & -2\Gamma_{10}\rho_{11} + 2\Gamma_{21}\rho_{22} & -(\Gamma_{10} + \Gamma_{21} + \gamma^{\varphi}_{21})\rho_{12} \\
        -(\Gamma_{21}+ \gamma^{\varphi}_{20})\rho_{20} & -(\Gamma_{10} + \Gamma_{21} + \gamma^{\varphi}_{21})\rho_{21} & -2\Gamma_{21}\rho_{22}
\end{array} \right] . \label{eq_liouville_rf_matrix}
\end{equation}
\end{widetext}

Another form for the dephasing Liouvillean used sometimes in the literature \cite{tsai}
can be obtained from Eq. (\ref{eq_liouville_rf}) by separating the diagonal and off-diagonal terms,
\begin{eqnarray}
{\cal L}^{\rm (eff)}[\rho] &=& \Gamma_{21}\rho_{22} (\sigma_{11} - \sigma_{22}) + \Gamma_{10} \rho_{11} (\sigma_{00} - \sigma_{11}) \nonumber \\
&& - \sum_{j,k\in\{0,1,2\};j\neq k} \frac{\gamma_{jk}}{2} \rho_{jk}\sigma_{jk} , \label{eq_liouville_rf2}
\end{eqnarray}
where now $\gamma_{10} = \gamma_{01} = \Gamma_{10} + \gamma_{10}^{\varphi}$, $\gamma_{20} = \gamma_{02} = \Gamma_{21} + \gamma_{20}^{\varphi}$, and $\gamma_{21} = \gamma_{12} = \Gamma_{21} + \Gamma_{10} + \gamma_{21}^{\varphi}$ have the meaning of off-diagonal decay rates for the corresponding density matrix elements. In general, as noted before, relaxation also results in dephasing, and in particular for three-level ladder systems such as the one discussed in this experiment
both $\Gamma_{10}$ and $\Gamma_{21}$ contribute to the the off-diagonal decay rate $\gamma_{21} = \gamma_{12}$.

Using now the master equation $\dot{\rho} = -(i/\hbar) [H^{\rm (eff)},\rho] + {\cal L}^{\rm (eff)}[\rho]$,
with the effective Hamiltonian Eq. (\ref{effective_hamiltonian}) and the effective Liouvillean Eq. (\ref{eq_liouville_rf}) we are able to provide a
complete characterization of the dynamics of the system.


\section{Spurious excitations to higher levels: a 5-level model}
\label{5levels}

In the previous sections we have developed a three-level model with dissipation for our phase qubit. Truncating the Hilbert space to three levels is a good approximation, as one can see when comparing the predictions of this model with the experimental data (next sections). In addition, as we have shown in the previous section, analytical results can be derived. However, due to the fact that the phase qubit is a multi-level system with finite anharmonicity, one source of errors for the three-level model comes from leakage to the higher excited levels. This is due to the off-resonant coupling of the fields with the higher-level transitions and subsequent transitions into the first three levels. This effect can be accounted for by a straightforward generalization of the three-level model to more levels, and then numerically solving the corresponding master equation. For our simulations we chose five levels. We found that further increasing the number of levels produces no significant change in the results. To avoid cumbersome equations, we only briefly describe the model here. We follow exactly the same procedure described in the previous two sections: starting from Eqs. (\ref{eq_hamiltonian2}) and Eqs. (\ref{eq_hamiltonian02}), we calculate the matrix elements of the perturbed harmonic oscillator $H' = H_0 + \frac{\Delta\Phi^3}{2\Phi_0 L^\ast}$ in the basis of $H_0$. The resulting $5\times 5$ matrix is diagonalized numerically; the time-dependent term $\Delta\Phi\Phi_{\rm rf}$ is then added and the full five-level time-dependent Hamiltonian $H(t)$ corresponding to Eq. (\ref{eq_hamiltonian5}) is obtained. For the dissipation, we use the straightforward generalization of the previous Liouvillean,
\begin{eqnarray}
{\cal L} [\rho_{\rm S}] &=& \sum_{j\in \{1,2,3,4\}}
\frac{\Gamma_{j,j-1}}{2}
(2\sigma_{j-1,j}\rho_{\rm S}\sigma_{j,j-1} - \sigma_{jj}\rho_{\rm S} -
\rho_{\rm S}\sigma_{jj})
\nonumber \\
&& - \sum_{j,k\in\{0,1,2,3,4\};j\neq k}\frac{\gamma^{\varphi}_{jk}}{2} \sigma_{jj}\rho_{\rm S} \sigma_{kk},\label{rell}
\end{eqnarray}
Note that in Eq. (\ref{rell}) we have neglected terms that mix the interlevel relaxation rates -- of the type appearing as the last one in Eq. (\ref{eq_relax}). The justification for this is that, similar to the approximation involved when going from Eq. (\ref{eq_liouville}) to Eq. (\ref{eq_liouville_rf}) in Section \ref{master_eq}, these terms will average out when moving to a multi-rotating frame (defined by extending Eq. (\ref{frame}) to five levels). Thus we can neglect them already in the Schr\"odinger picture.

Finally, we put the master equation $(d/dt)\rho_{\rm S}(t) = -(i / \hbar)\left[ H(t), \rho_{\rm S}(t) \right] + {\cal L}[\rho_{\rm S}(t)]$ in matrix form and solve it numerically.


\section{Determination of the dephasing rates}
\label{dephasing}

\begin{figure}[h]
\includegraphics[width=8cm]{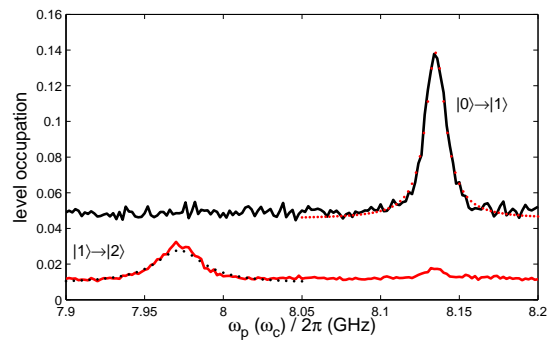}
\caption{(color online). The spectroscopy traces (solid curves) taken from Fig. 1(c) in Ref. [\onlinecite{mika}]. The (black) solid trace is the single-tone spectroscopy experiment on the effective two-level system showing the $|0\rangle\rightarrow|1\rangle$ spectral line centered at 8.135 GHz.  The (red) solid trace is a two-tone spectroscopy, revealing the $|1\rangle\rightarrow|2\rangle$ spectral line centered at 7.975 GHz. The dotted curves are fits to simulations of the full five-level master equation. The value of the probe field Rabi frequency used was $\Omega_{p} = 2\pi \times 3.5 $ MHz.} \label{fig_spectro}
\end{figure}

We now address the question of how to determine the numerical values of the parameters that enter in the Hamiltonian and in the Liouvillean of the models presented. First, a number of independent experiments are performed that allow us to find $\omega_{10},\omega_{21}, \Gamma_{10}$, and $\Gamma_{21}$. The transition frequencies $\omega_{10} = 2\pi \times 8.125$ GHz and $\omega_{21} = 2 \pi \times 7.975$ GHz, are determined immediately from standard single-tone and two-tone spectroscopy (see Fig. \ref{fig_spectro}). The relaxation rates are determined as follows: for $\Gamma_{10}$ we apply a $\pi$ pulse resonant to the first transition, taking the system from $|0\rangle$ to $|1\rangle$. Then we let the qubit decay and determine the occupation probability by delaying the measurement pulse with respect to the $\pi$ pulse. This probability decays exponentially as a function of time, allowing us to extract the relaxation $\Gamma_{10}=2\pi\times 7$ MHz. The same technique can be applied to record the decay of the $|2\rangle$ state to the $|1\rangle$ state, while simultaneously driving the first transition continuously to ensure a non-zero population on the state $|1\rangle$, and we get $\Gamma_{21} = 2\pi \times 11$ MHz.

The  pure dephasing rates $\gamma^{\varphi}_{10}$, $\gamma^{\varphi}_{20}$, and $\gamma^{\varphi}_{21}$  can be obtained from the spectroscopic linewidths in the following way. We first do a single-tone spectroscopy, measuring the sum of the occupation probabilities of the excited states at zero coupling field (black solid curve in Fig. \ref{fig_spectro}). Since $\Omega_c = 0$ in the first approximation we may treat the system as a quasi-two-level system under monochromatic driving, and neglect the leakage to higher levels. This experiment thus represents a simple spectroscopy of the two-level system $\{|0\rangle, |1\rangle\}$.
We, therefore, can write the master equation with the rotating-frame effective Hamiltonian Eq. (\ref{effective_hamiltonian}) and
Liouvillean Eq. (\ref{eq_liouville_rf_matrix}) truncated to the ground state and first excited state and obtain
\begin{widetext}
\begin{equation}
\dot{\rho} = \frac{1}{2}\left[ \begin{array}{cc}
        i\Omega_p(\rho_{01}-\rho_{10}) + 2\Gamma_{10}\rho_{11} & -i[\Omega_p(\rho_{11}-\rho_{00}) - 2\Delta_p\rho_{01}] - \gamma_{10}\rho_{01} \\
        -i[\Omega_p(\rho_{00}-\rho_{11}) + 2\Delta_p\rho_{10}] - \gamma_{10}\rho_{10} & -i\Omega_p(\rho_{01}-\rho_{10}) - 2\Gamma_{10}\rho_{11}
\end{array} \right] , \label{eq_master_two_level}
\end{equation}
\end{widetext}
where we use the same notations as before,  $\Delta_p = \omega_{10} - \omega_p$, and $ \gamma_{10} = \Gamma_{10} + \gamma^{\varphi}_{10}$. In the steady state $\dot{\rho} = 0$, and the occupation probability of level $|1\rangle$, as a function of detuning, is obtained as
\begin{equation}
P_1(\Delta_p) = \frac{\Omega_p^2\gamma_{10}}{4\Delta_p^2\Gamma_{10} + \gamma_{10}\left[ 2\Omega_p^2 + \Gamma_{10}\gamma_{10} \right]} .
\end{equation}
This probability reaches its maximum value at $\omega_p = \omega_{10}$.

The width $\delta f_{10}=\delta \omega_{10}/2\pi$ of this spectroscopy peak,
\begin{equation}
\delta \omega_{10} = 2\pi\delta f_{10} = \sqrt{\gamma_{10}^2 + 2\Omega_p^2 \gamma_{10}/\Gamma_{10}} \approx \gamma_{10} , \label{eq_peak_width}
\end{equation}
is determined by solving the equation $P_1(\Delta_p = \pi \delta f_{10}) = P_1(\Delta_p=0) / 2$.
The  result shows, as expected, that for relatively small probe fields the width of a spectroscopy line of a two-level system is given by the sum of the relaxation rate and the pure dephasing rate. Using Eq. (\ref{eq_peak_width}) we can now determine $\gamma^{\varphi}_{10}$. We use the values of the independently-measured decay rate $\Gamma_{10} = 2\pi\times 7$ MHz and find $\delta f_{10} =  14$ MHz from the black continuous line (the measured data) in Fig.  \ref{fig_spectro}. This yields $\gamma^{\varphi}_{10} \approx \delta \omega_{10} - \Gamma_{10} =  2\pi\times 7$ MHz.

To obtain the other dephasing rates $\gamma^{\varphi}_{20}$ and $\gamma^{\varphi}_{21}$ there is no simple analytical result that we can use, therefore we need to rely on simulations and direct fittings with spectroscopy traces involving level $|2\rangle$. We write $\gamma^{\varphi}_{21} =  \gamma^\varphi_{10}  +  \gamma^\varphi_{20} - \varepsilon$, and we find that the best fittings are obtained for $\gamma^{\varphi}_{20} = 2\pi\times 16$ MHz and $\varepsilon = 2\pi\times 5$ MHz ($\gamma^{\varphi}_{21} = 2\pi \times 18$ MHz). As we show in Appendix \ref{appendixb}, $\varepsilon$ is a measure of correlation between the fluctuations of level 1 and level 2. Reasonably good fittings are obtained also for $\varepsilon = 0$ MHz (see Fig. (\ref{fig_spectro})). As we will see in the next section, the advantage of the approximation
$\varepsilon \approx 0$ MHz is that relatively simple analytical results can be obtained.

In Table \ref{table} we summarize our results for the parameters of the three-level model. Even better fittings can be obtained if one employs the five-level model described in Section \ref{5levels}, thus accounting for the leakage outside the subspace spanned by $\{|0\rangle ,|1\rangle , |2\rangle \}$.  For this subspace we now have all the numerical values needed to simulate the dynamics. Now, in the Hamiltonian, the additional matrix elements corresponding to the fourth and fifth states ($|3\rangle $ and $|4\rangle $) can be calculated based on the information already obtained from the three-level model. The only unknowns are the dissipation parameters associated with the last two levels, for which an independent experimental determination is not easily available.  A rough estimation of these parameters is based on the result that, since dissipation is due to coupling to the external electromagnetic environment, we can expect the same scaling to continue. This estimation does not need to be too precise: we have checked numerically that even relatively large errors in the estimation of these higher-level dissipation parameters do not result in significant mismatches with the experimental data. The results of simulating the five-level model are plotted with dotted lines in Fig. \ref{fig_spectro}. The dotted red curve fits almost perfectly the $|0\rangle\leftrightarrow|1\rangle$ transition, and a very good fit (black dotted lines) is obtained for the $|1\rangle\leftrightarrow|2\rangle$ transition. Thus, we conclude that the five-level model confirms the predictions of  the three-level model and further improves the fitting with the experimental data.


\section{Analytical results for three level systems}
\label{analytical}

In this section we show that relatively simple analytical results can be obtained if we use the approximation
$\gamma_{21} \approx \gamma_{10} + \gamma_{20}$, where
$\gamma_{10} = \Gamma_{10} + \gamma^{\varphi}_{10}$, and $\gamma_{20} = \Gamma_{21} + \gamma^{\varphi}_{20}$
are the total off-diagonal decay rates introduced before in Eq. (\ref{eq_liouville_rf2}). This relation is satisfied reasonably well for our system, as
we show in Appendix \ref{appendixb}. Note that although there is no {\it a priori} reason for the general validity of this equation in a three-level system, we do expect it to be a good approximation for many cases of interest. Indeed, the error that we make by this approximation in the total off-diagonal decay rate $\gamma_{21}$
is $\varepsilon = \gamma_{10}^{\varphi} + \gamma_{20}^{\varphi} - \gamma_{21}^{\varphi}$, which should be compared with $\gamma_{10} + \gamma_{20}$. But the latter quantity is relatively large, since it contains the sum of two relaxation rates and two pure dephasing rates, $\gamma_{10} + \gamma_{20} = \Gamma_{21} +\Gamma_{10} + \gamma_{20}^{\varphi} + \gamma_{10}^{\varphi}$. For example, in our case we have $\gamma_{10} + \gamma_{20} = 2\pi \times 41$ MHz and $\varepsilon = 2\pi \times 5$ MHz, meaning that our approximation results in an overestimation of about 13\%  of the true value of $\gamma_{21}$.


Using  Eqs. (\ref{effective_hamiltonian}) and (\ref{eq_liouville_rf}) we can now solve  for the density matrix $\rho$, which  satisfies $\dot{\rho} = -(i/\hbar)\left[ H^{\rm (eff)}, \rho \right] + {\cal L}^{\rm (eff)}[\rho]$. In the steady state, $\dot{\rho}^{\rm (st)}=0$, and the elements of the density matrix can be found in analytical form using  Mathematica. The full solution can be put in a simpler form if some further approximations (see also Appendix \ref{appendixc}), corresponding to the actual values used in the experiment, are used.

The first approximation that we employ is that the ratio between $\Omega_{p}^{2}$ and any product of two $\gamma$'s is much smaller than 1 ($\Omega_{p}^{2}\ll \gamma_{20}^2, \gamma_{10}^2, \gamma_{10}\gamma_{20}$); the largest ratio for our experiment is $\Omega_{p}^{2}/\gamma_{10}^2=0.06$. Next, we neglect the square of the  probe field with respect to the square of  the coupling field ($\Omega_p^{2}/\Omega_c^{2}\ll 1$); for the parameters used in our experiment, this ratio is smaller than 0.02.

Under these conditions, we obtain the stationary occupation probability of level $|1\rangle$,
\begin{widetext}
\begin{equation}
P_1 = \rho_{11}^{\rm (st)} \approx \frac{\Omega_p^2 \left[4\Delta_p^2\gamma_{10} + \gamma_{20}\left( \Omega_c^2 + \gamma_{10}\gamma_{20} \right) \right]}{\Gamma_{10}\left[ 16\Delta_p^4 + \left( \Omega_c^2 + \gamma_{10}\gamma_{20} \right)^2 + 4\Delta_p^2 \left( \gamma_{10}^2 + \gamma_{20}^2 - 2\Omega_c^2 \right) \right]} . \label{eq_occupation_prob}
\end{equation}
\end{widetext}

This analytical equation agrees well with the experimental data and also with a more complete five-level model described previously. For $\Delta_{p}=0$, {\it i.e.} at the frequency where in the absence of the coupling field there was an absorption peak, Eq. (\ref{eq_occupation_prob}) yields $P_{1}  = \gamma_{20}\Gamma_{10}^{-1}\Omega_{p}^2(\Omega_{c}^2 + \gamma_{10}\gamma_{20})^{-1}$, showing that the population of level 1 decreases to zero as $\Omega_{c}$ increases. As a function of $\Delta_p$ (or $\omega_p$) the occupation probability has a double-peak shape (see later in Fig. (\ref{fig_Autler}). We will refer to these peaks as the Autler-Townes peaks. A more detailed analysis is presented in Section \ref{comparison}.

The splitting $\delta f_{\rm AT}=\delta \omega_{\rm AT}/2\pi$ is twice the value of the positive solution of the equation $\partial_{\Delta_p}P_1 = 0$; we get
\begin{widetext}
\begin{equation}
\delta \omega_{\rm AT} = 2\pi \delta f_{\rm AT} = \sqrt{\frac{\Omega_c (\gamma_{10}+\gamma_{20})\sqrt{\Omega_c^2 + \gamma_{10}\gamma_{20}} - \gamma_{20}\left( \Omega_c^2 + \gamma_{10}\gamma_{20} \right)}{\gamma_{10}}} . \label{eq_splitting}
\end{equation}
\end{widetext}
When comparing this formula with the experimental data the agreement is very good (see Section \ref{comparison}, Fig. \ref{splt}). One also notes that if $\Omega_{c}$ is much larger than the decoherence rates the dependence is linear  $\delta \omega_{\rm AT}\approx\Omega_c$; otherwise, dissipation tends to move the Autler-Townes splitting away from linearity.

The matrix elements $\rho_{01}=\rho_{10}^{*}$ can be calculated under the same approximation of a weak probe field as above, with the result
\begin{equation}
\rho_{10} =\frac{\Omega_{p}(2\Delta_{p} - i\gamma_{20})}{-4\Delta_{p}^{2}+\Omega_{c}^{2} + \gamma_{10}\gamma_{20}+ 2i\Delta_{p}(\gamma_{10}+\gamma_{20})}.\label{rho10}
\end{equation}
Although this off-diagonal matrix element is not directly measurable in our experiment, it plays an important role in other experimental configurations \cite{tsai,astafiev}.

The rest  of the elements of the density matrix are too complicated to be listed here. Still, we can make progress by restricting ourselves to the case of coupling fields larger then all the linewidths (more precisely, the case in which the products $\gamma_{20}^2$, $\gamma_{10}^2$ and $\gamma_{10}\gamma_{20}$ can be neglected with respect to $\Omega_{c}$). The expressions for the elements of the density matrix in this case are given in Appendix \ref{appendixc}.


\section{Comparison with the Autler-Townes experiment}
\label{comparison}

\begin{figure}[h]
\includegraphics[width=8.5cm]{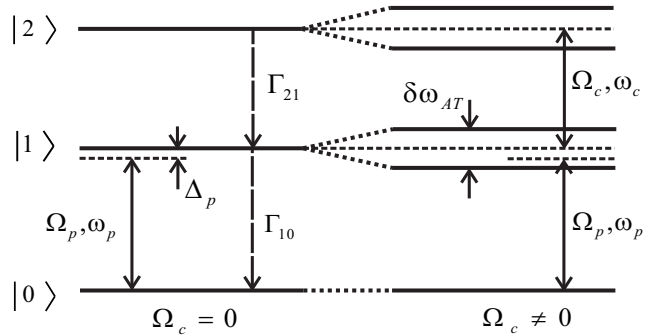}
\caption{Schematic of the Autler-Townes effect as a consequence of the formation of dressed states for the $|2\rangle\leftrightarrow|1\rangle$ transition due to a resonant coupling field.} \label{ATscheme}
\end{figure}

In the previous sections we have described how to obtain the parameters entering in the Hamiltonian and in the Liouvillean, which determine the dynamics of the qubit (for the three-level model, see Table \ref{table}). In this section we describe how to use this information for modeling the behavior of the system at large coupling powers, which produces the Autler-Townes effect. This effect is seen in the spectroscopy trace as the splitting (sometimes called dynamic Stark splitting) of the line corresponding to the first transition into two peaks. In the doubly-rotating frame, this can be understood as a consequence of the dressing of the transition $|1\rangle\rightarrow|2\rangle$ by the electromagnetic field $\Omega_c$ to which it is coupled (Fig. \ref{ATscheme}). Effectively, the weak probe field $\Omega_p$ sees the doublet of the two dressed states formed around the qubit state $|1\rangle$. The dressed-state picture follows naturally when the fields are treated quantum-mechanically\cite{Zubairy}, but a similar picture emerges also in our formalism, which uses classical fields. From Eq. (\ref{eq_liouville_rf}), by neglecting the probe field and with the coupling field at resonance $\omega_{c}=\omega_{21}$, the two nonzero eigenvalues form a doublet separated by an energy $\hbar\Omega_c$ (which is $\delta\omega_{\rm AT}$ at $\Omega_{p}=0$).

\begin{table}
\begin{tabular}{||l||c|c|c||}\hline\hline
     parameter & $\omega_{i,i-1}$ (GHz) & $\Gamma_{i,i-1}$ (MHz) & $\gamma_{i0}^{\varphi}$ (MHz)\\ \hline \hline
$i=1$ &  $2\pi \times 8.135$   & $2\pi \times 7$   &  $2\pi \times 7$    \\ \hline
$i=2$ & $2\pi \times 7.975$      &  $2\pi \times 11$  & $2\pi \times 16$ \\ \hline\hline
\end{tabular}
\caption{Parameters for the three-level model. For $\gamma^{\varphi}_{21}$ we find  $\gamma^{\varphi}_{21} = 2\pi \times 18$ MHz.}
 \label{table}
\end{table}

\begin{figure}[h]
\includegraphics[width=8.5cm]{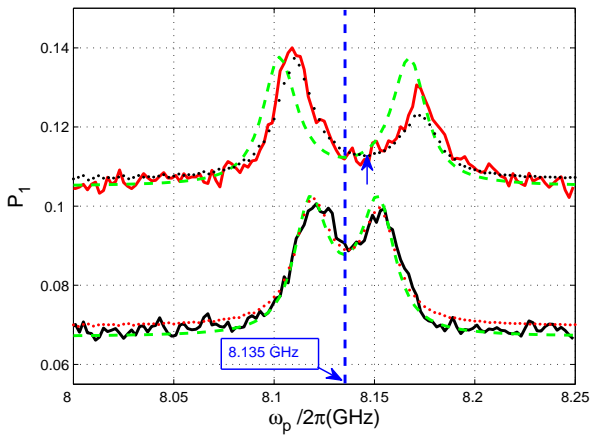}
\caption{(color online). Autler-Townes splitting in a phase qubit. The experimental spectroscopy traces (solid lines) are taken from Fig. 3(a) in Ref. [\onlinecite{mika}]. The dotted lines are simulations of the master equation corresponding to the full five-level model, with the parameters previously extracted from the spectroscopy data. The (green) dashed lines are plotted by using Eq. (\ref{eq_occupation_prob}) with the same values (the offsets of base level for $\Omega_c = 2\pi\times 36$ MHz and $\Omega_c = 2\pi\times 66$ MHz are 0.057 and 0.105, respectively). Note that the inclusion of higher levels at large values of the coupling results in asymmetric Autler-Townes peaks and in the displacement of the spectroscopy curve toward higher frequencies. The vertical (blue) dashed line indicates the first transition $\omega_{10}=2\pi \times 8.135$ GHz. At this frequency the lower curve (taken at $\Omega_c = 2\pi\times 36$ MHz) has a minimum, while for the higher curve (taken at $\Omega_c = 2\pi\times 66$ MHz) the minimum, marked by the blue vertical arrow, is displaced by 10 MHz to a value of 8.145 GHz. } \label{fig_Autler}
\end{figure}

In Fig. \ref{fig_Autler} we show the same data as in Fig. 3(a) in Ref. [\onlinecite{mika}]. The (black) solid curve corresponds to $\Omega_c = 2\pi\times 36$ MHz and the red one to $\Omega_c = 2\pi\times 66$ MHz; the value of $\Omega_p$ was $2\pi \times 3$ MHz. We first fit the experimental data by using the analytical model with three levels presented in Section \ref{master_eq} (see Eq. (\ref{eq_occupation_prob})). The results are shown with (green) dashed lines in Fig. \ref{fig_Autler}. Then we use the numerical solution of the full five-level model to obtain the dotted lines in Fig. \ref{fig_Autler}, and one sees that the fit is almost perfect. When plotting all the numerical results one has to remember to add the residual tunneling amplitude of approximately 0.057,  which was mentioned in Ref. [\onlinecite{mika}], and which is also visible in the off-resonant part of the black trace in Fig. \ref{fig_spectro}. This is specific to the measuring scheme (based on switching of a nearby SQUID). This offset also has to be added when comparing Eq. (\ref{eq_occupation_prob}) with the measured data. At high coupling powers, the field $\Omega_c$ will couple to the $|0\rangle \rightarrow |1\rangle$ transition as well, an effect that is not captured by Eq. (\ref{eq_occupation_prob}). As a result, when using Eq. (\ref{eq_occupation_prob}), we expect that at low powers the value of this offset is close to 0.057, but at high powers (e.g. $\Omega_c = 2\pi\times 66$ MHz) the offset value has to be adjusted appropriately (see Fig. \ref{fig_Autler}).

\begin{figure}[h]
\includegraphics[width=7cm]{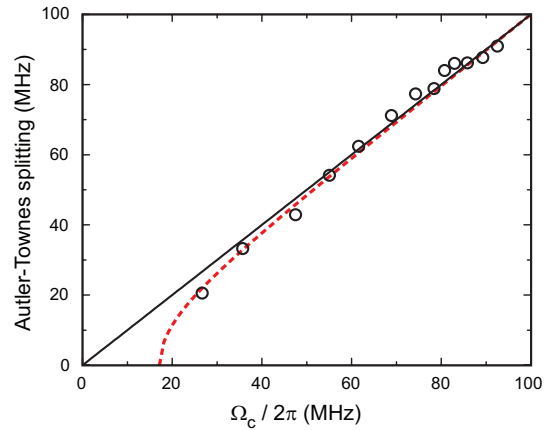}
\caption{(color online). The Autler-Townes frequency splitting $\delta\omega_{\rm AT}/(2\pi )$ as given by the simplified three-level model with dissipation Eq. (\ref{eq_splitting}), shown as a (red) dotted line. The straight line is the prediction of the three-level model without dissipation. The experimental data (circles) are from Fig. 3(c) of Ref. [\onlinecite{mika}].} \label{splt}
\end{figure}

Another observation regarding the three-level model is that it predicts symmetric Autler-Townes peaks, while the experiment and the full model give peaks with slightly different heights and displaced to higher frequencies (see Fig. \ref{fig_Autler}). However, the distance between these peaks (the Autler-Townes splitting) is predicted remarkably well in this simple model. Indeed, in Fig. \ref{splt} we show a comparison between experimental data (as in Fig. 3(c) in Ref. [\onlinecite{mika}]) and the results for splitting as given by Eq. (\ref{eq_splitting}). We notice that the three-level model with dissipation predicts a bending of the Autler-Townes frequency splitting at low powers, which is observed in the experiment and is not captured by the analysis of the energy levels in the three-level model without dissipation.

Finally, we can simulate the case when both the coupling and probe frequencies are swept, and we get again a very good agreement with the data, as in  Fig. 4 of Ref. [\onlinecite{mika}].


\section{Creation of dark states}
\label{dark}

\begin{figure}[h]
\includegraphics[width=9.5cm]{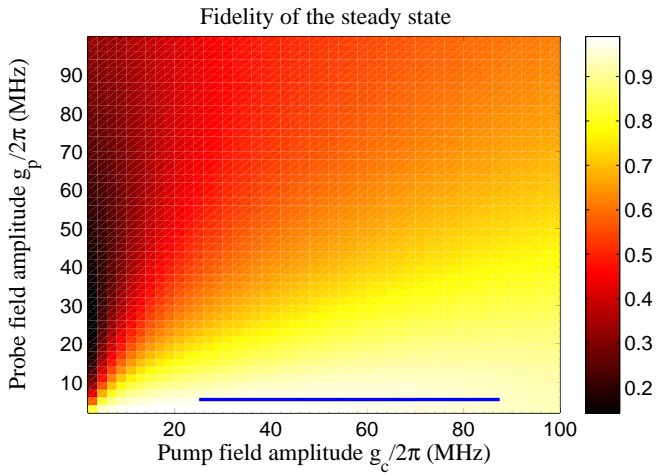}
\caption{The fidelity of the steady-state as a function of the pump and probe amplitudes. The probe and field amplitudes are directly related to the respective Rabi frequencies by $\Omega_{p} = 0.69g_{p}$ and $\Omega_{c}=g_{c}$. The blue line corresponds to $\Omega_{p}=2\pi \times 3.5$ MHz ($g_{p}=2\pi \times 5$ MHz), and to the range of experimental values  for $\Omega_{c}$ presented in Fig. \ref{splt}.} \label{fidelity}
\end{figure}

\begin{figure}[h]
\includegraphics[width=9.5cm]{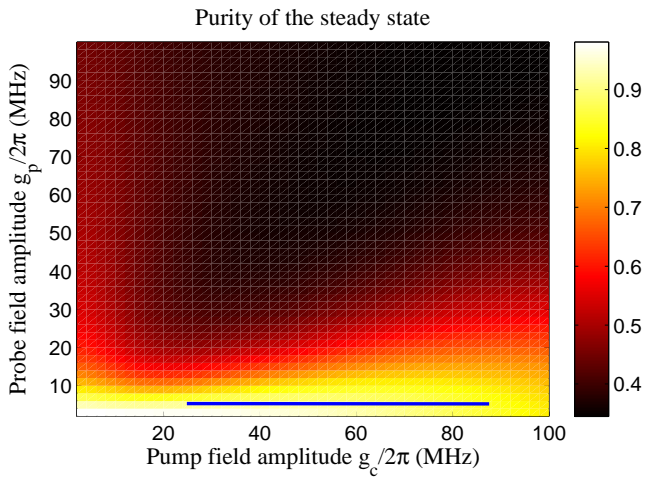}
\caption{The purity of the steady-state as a  function of the pump and probe amplitudes. As in Fig. \ref{fidelity}, the blue line corresponds to $\Omega_{p}=2\pi \times 3.5$ MHz ($g_{p}=2\pi \times 5$ MHz) and a range of experimental values  for $\Omega_{c}$ as presented in Fig. \ref{splt}. Also, the probe and field amplitudes are directly related to the respective Rabi frequencies by $\Omega_{p} = 0.69g_{p}$ and $\Omega_{c}=g_{c}$.} \label{purity}
\end{figure}

An important fundamental issue is whether true dark states and electromagnetically induced transparency can be created in systems of superconducting qubits. A  dark state is the zero-eigenvalue eigenvector of the effective Hamiltonian $H^{(\rm eff)}$ of Eq. (\ref{eq_liouville_rf}) for the resonant case, $\Delta_{p} = \Delta_{c} = 0$. The Hamiltonian is diagonalized as
\begin{eqnarray}
&|\rm D\rangle = \cos \Theta |0\rangle - \sin \Theta |2\rangle& {\rm eigenval.=} 0 ,\nonumber \\
&\frac{1}{\sqrt{2}}\left[\sin \Theta |0\rangle \pm  |1\rangle +\cos \Theta |2\rangle\right]&
{\rm eigenval.=}\pm\sqrt{\Omega_{p}^{2} + \Omega_{c}^2},\nonumber
\end{eqnarray}
where the mixing angle $\Theta$ is given by $\tan \Theta = \Omega_{p}/\Omega_{c}$. The first state is characterized by the absence of population on $|1\rangle$; contrary to this, the other two eigenvalues have a considerably larger population of the state $|1\rangle$. Moreover, in the limit $\Omega_{p}\ll \Omega_{c}$ (which corresponds well to our experimental parameters), we notice that the dark state becomes approximately equal to the ground state $|0\rangle$, and the other two eigenvectors become approximately $(1/\sqrt{2})(\pm|1\rangle + |2\rangle )$. Thus, ground-state amplitudes are 1 for the dark state and 0 for the other two eigenstates. These features are a hint that the dark state is easily distinguishable from  the other two eigenstates. Indeed, by looking at the experimental data presented above (small populations for the levels 1 and 2) it is clear that they are consistent with the stationary state of the system being close to a dark state. In order to quantify this observation, we calculate the distance between the steady-state $\rho^{\rm (st)}$ and the dark state $|D\rangle$ by the fidelity \cite{nielsen}
\begin{eqnarray}
{\cal F}_{|\rm D\rangle}[\rho^{\rm (st)}] &=& \sqrt{\langle \rm D|\rho^{\rm (st)}|\rm D\rangle} \\
 &=&\frac{\cos 2\Theta}{2}(\rho_{00}^{(\rm st)}-\rho_{22}^{(\rm st)}) -
\frac{\sin 2\Theta}{2}(\rho_{20}^{(\rm st)}+\rho_{02}^{(\rm st)}) \nonumber \\
&& +\frac{1}{2}(1-\rho_{11}^{(\rm st)}),\label{fid}
\end{eqnarray}
using  $\Omega_{p}$ and $\Omega_{c}$ as adjustable parameters. The results are presented in Fig. (\ref{fidelity}). We have also checked that the same result is obtained in the Schr\"odinger picture (the non-rotating frame) by evolving numerically the system for a time much longer than all the timescales set by dissipation and driving, and calculating the fidelity with respect to the dark state in the non-rotating frame, where  it takes the form $\cos\Theta |0\rangle + \exp [-i(\omega_c+\omega_p)t]\sin\Theta|2\rangle$. Similarly, one can calculate the state purity ${\rm Tr}\{\left(\rho^{(\rm st)}\right)^2\}$; the result is plotted in Fig. \ref{purity}. As expected, for relatively low amplitudes of the probe pulse, as used in our experiments, the state is very close to a pure state.

For our experiment the fidelity reached values above 0.95 (see Fig. \ref{fidelity}) at pumping fields $\Omega_{c}$ higher than $2\pi\times 50$ MHz. Thus, to a reasonable approximation, these experiments can create steady states that are close to ideal dark states. Steady states could be useful in future quantum processors for state preparation and as sources of entanglement\cite{njp}. A more in-depth characterization of these states, and the effect of relaxation and dephasing, is given in Appendix \ref{appendixb}. One can anticipate, for example, that the decay from the level $|2\rangle$ will inevitably decrease its population, and will "contaminate" the true dark state with an incoherent population on level $|1\rangle$. This observation is confirmed by the analysis in Appendix \ref{appendixc}. A somewhat more advantageous situation could be achieved if the system is used in a $\Lambda$-configuration, in which case the lifetime of the upper-state component of the dark state is increased \cite{bbn}. Yet, even for the ladder configuration discussed here, the stationary state is close to an ideal dark state.

In general, the main obstacles in increasing the fidelity are the relatively short (compared to atomic physics) decoherence times. However, the advantage offered by artificial systems is that larger couplings with the fields can be realized: this allows experimentalists to increase the strength of the field until the Autler-Townes splitting well exceeds the natural spectral linewidth.


\section{Conclusions}
\label{conclusion}

We have studied a phase qubit under a two-tone microwave irradiation applied close to the $|0\rangle \rightarrow |1\rangle$ and $|1\rangle \rightarrow |2\rangle$ transition frequencies. We provided a detailed quantum-mechanical theoretical description of the three-level system interacting with the fields, including relaxation and dephasing. The relaxation and dephasing rates were extracted from independent experiments. These parameters were then  used in the analysis of the Autler-Townes effect to compare the measurements with the theoretical predictions. We found that the value of the Autler-Townes splitting and the occupation probabilities predicted by the three-level model are in very good agreement with the experimental data. This agreement improves if the next higher levels are included and full numerical simulations are performed. Finally, we analyzed the structure of the stationary states thus created, and show that they are very close to true dark states.


\acknowledgements
G.S.P. thanks Petr Anisimov for useful discussions. We acknowledge financial support from the Academy of Finland (no. 129896, 118122, 130058, and 135135), from the National Graduate School of Material Physics, from NIST, and from the European Research Council (StG).

\appendix


\section{Energy relaxation}
\label{appendixa}

We define the density matrix of our whole system ($system\ S \oplus reservoir\ R$) as $\chi(t)$. Since our interest centers only around the resulting state of the system $S$, irrespective of the outcome of $R$, we want to obtain an equation of the reduced density matrix $\rho(t) = {\mathrm{Tr}}_R [\chi(t)]$ in order to describe the dynamics of the system $S$ induced by the evolution of the total system. The aim is to derive a so-called {\em Markovian master equation} for the reduced density matrix $\rho$ following Carmichael \cite{Carmichael}, and Breuer and Petruccione \cite{Breuer}.

The Hamiltonian of the whole system is assumed to have the general form
\begin{equation}
H_{tot} = H_S + H_R + H_{SR} , \label{A1}
\end{equation}
where
\begin{equation}
H_S = \hbar \omega_{10}|1\rangle\langle 1| + \hbar \left(\omega_{10} + \omega_{21}\right)|2\rangle\langle 2| \label{A2}
\end{equation}
denotes the Hamiltonian of a three-level system in its energy eigenbasis $\{|0\rangle, |1\rangle ,|2\rangle \}$, with transition frequencies $\omega_{10}$, and
$\omega_{21}$. The Hamiltonian $H_R$,
\begin{equation}
	H_R = \hbar\sum_\alpha\omega_\alpha a_\alpha^\dagger a_\alpha \label{A3}
\end{equation}
is a collection of harmonic oscillators with frequency $\omega_\alpha$ and corresponding creation (annihilation) operator $a_\alpha^\dagger$ ($a_\alpha$), and
\begin{equation}
H_{SR} = \hbar\sum_\alpha \left[ a_\alpha^\dagger \left( f_{1\alpha}|0\rangle\langle 1| + f_{2\alpha}|1\rangle\langle 2| \right) + h.c. \right] \label{A4}
\end{equation}
describes the coupling between the inter-level transitions of the three-level system and the reservoir oscillators. The coupling constants $f_{j\alpha}$ ($j=1,2$) are assumed to be small (weak-coupling limit). Note that we do not take the $|0\rangle \leftrightarrow |2\rangle$ transition into account due to the small anharmonicity of the system.

We transform the Schr{\"{o}}dinger equation for the density operator of the whole system $\chi$, (from now on we take $\hbar = 1$)
\begin{equation}
\dot{\chi} = -i[H_{tot},\chi] , \label{A5}
\end{equation}
into the interaction picture, with $\tilde{\chi}(t) = \exp[i(H_S + H_R)t] \chi (t) \exp[-i(H_S + H_R)t]$, and obtain
\begin{equation}
\dot{\tilde{\chi}} = -i [ \tilde{H}_{SR}(t),\tilde{\chi} ] , \label{A6}
\end{equation}
where $\tilde{H}_{SR}(t)$ is given by $\tilde{H}_{SR}(t) = \exp[i(H_S + H_R)t] H_{SR} \exp[-i(H_S + H_R)t]$.

By integrating Eq. (\ref{A6}) and substituting for $\tilde{\chi}(t)$ inside the commutator in equation (\ref{A6}), we get a Liouville equation in the integro-differential form
\begin{eqnarray}
	\dot{\tilde{\chi}}(t) &=& -i [ \tilde{H}_{SR}(t),\chi (0) ]  \nonumber \\
& & -\int_0^t dt' \left[ \tilde{H}_{SR}(t), [ \tilde{H}_{SR}(t'),\tilde{\chi}(t') ] \right] . \label{A7}
\end{eqnarray}
The reduced density matrix for the system $S$ in the interaction picture is given by
\begin{equation}
	\tilde{\rho}(t) = \mathrm{Tr}_R \tilde{\chi}(t) = e^{iH_St}\rho(t)e^{-iH_St} . \label{A7.1}
\end{equation}
We assume that the interaction between $S$ and $R$ is switched on at time $t = 0$. So at this initial time, $S$ and $R$ are uncorrelated and the total density matrix is factorized as the direct product
\begin{equation}
	\chi(0) = \rho(0)\otimes R(0) = \tilde{\chi}(0) , \label{A7.2}
\end{equation}
where $R(0)$ denotes the initial reduced density matrix for the reservoir.

In order to eliminate $\chi$ from Eq. (\ref{A7}), we perform the first approximation, known as {\em Born approximation}, which assumes that the influence of the system on the reservoir is negligibly small due to the weak coupling between them. Therefore, the total density matrix $\tilde{\chi}(t)$ at time $t$ can be expressed as a tensor product
\begin{equation}
	\tilde{\chi}(t) = \tilde{\rho}(t)\otimes R(0) , \label{A7.3}
\end{equation}
with $R(0) = \exp(-\beta H_R) / Z$, where $\beta = 1 / k_BT$, and the partition function is $Z = \mathrm{Tr}[\exp(-\beta H_R)]$.

Hence, tracing both sides of Eq. (\ref{A7}) over the reservoir variables, and eliminating the term $-i {\mathrm{Tr}}_R [ \tilde{H}_{SR}(t),\chi (0) ]$ with the assumption
${\mathrm{Tr}}_R [ \tilde{H}_{SR}(t)R(0) ]=0$, we write Eq. (\ref{A7}) as
\begin{equation}
	\dot{\tilde{\rho}}(t) = - \int_0^t dt' {\mathrm{Tr}}_R \left\{ \left[ \tilde{H}_{SR}(t),[ \tilde{H}_{SR}(t'),\tilde{\rho}(t')\otimes R(0) ] \right] \right\} . \label{A8}
\end{equation}
In general, the evolution of the system depends on its past history ($\tilde{\rho}(t')$ in the integral). However, it is known that for a reservoir with many degrees of freedom this memory can be erased ({\em Markov approximation}), which allows us to make the replacement
\begin{equation}
	\tilde{\rho}(t') \rightarrow \tilde{\rho}(t) , \label{A9}
\end{equation}
and leads us the following master equation
\begin{equation}
	\dot{\tilde{\rho}}(t) = - \int_0^t dt' {\mathrm{Tr}}_R \left\{ \left[ \tilde{H}_{SR}(t),[ \tilde{H}_{SR}(t'),\tilde{\rho}(t)\otimes R(0) ] \right] \right\} . \label{A10}
\end{equation}

To get a more explicit form for the master equation, we need to insert $\tilde{H}_{SR}$ into Eq. (\ref{A10}), so we rewrite Eq. (\ref{A4}) as
\begin{equation}
	H_{SR} = \sum_{j=1,2} \left( \sigma_{j-1,j} \Gamma_j^\dagger + h.c. \right) , \label{A11}
\end{equation}
where $\sigma_{j,k} \equiv \sigma_{jk} \equiv |j\rangle\langle k|$ are operators for the system and $\Gamma_j^\dagger \equiv \sum_\alpha f_{j\alpha} a_\alpha^\dagger$ are operators for the reservoir. Transforming into the interaction picture by $\tilde{\sigma}_{j-1,j}(t) = \exp (iH_St)\sigma_{j-1,j}\exp (-iH_St)$ and $\tilde{\Gamma}_j^\dagger(t) = \exp (iH_Rt)\Gamma_j^\dagger \exp (-iH_Rt)$, all these operators become
\begin{eqnarray}
	&& \tilde{\sigma}_{j-1,j}(t) = \sigma_{j-1,j}e^{-i\omega_{j,j-1}t} , \nonumber \\
	&& \tilde{\sigma}_{j-1,j}^\dagger(t) = \tilde{\sigma}_{j,j-1}(t) = \sigma_{j,j-1}e^{i\omega_{j,j-1}t} , \nonumber
\end{eqnarray}
and
\begin{eqnarray}
	\tilde{\Gamma}_j^\dagger(t) = \sum_{\alpha} f_{j\alpha}a_\alpha^\dagger e^{i\omega_\alpha t} , \ \ \ \ \tilde{\Gamma}_j(t) = \sum_{\alpha} f_{j\alpha}^\ast a_\alpha e^{-i\omega_\alpha t} . \nonumber
\end{eqnarray}

The resulting master equation in the {\em Born-Markov approximation} is then
\begin{widetext}
\begin{eqnarray}
	\dot{\tilde{\rho}}&=& - \sum_{j,k = 1,2} \int_0^t dt' {\mathrm{Tr}}_R \left\{ \left[ \tilde{\sigma}_{j-1,j}(t)\tilde{\Gamma}_j^\dagger(t) + h.c.,[ \tilde{\sigma}_{k-1,k}(t')\tilde{\Gamma}_k^\dagger(t') + h.c.,\tilde{\rho}\otimes R(0) ] \right] \right\} \nonumber \\
	&=& \sum_{j,k = 1,2} \int_0^t dt' \left\{ \left[ \tilde{\sigma}_{k-1,k}(t') \tilde{\rho} \tilde{\sigma}_{j-1,j}(t) - \tilde{\sigma}_{j-1,j}(t) \tilde{\sigma}_{k-1,k}(t') \tilde{\rho} \right] \langle \tilde{\Gamma}_j^\dagger(t) \tilde{\Gamma}_k^\dagger(t') \rangle_R \right. \nonumber \\
	&& \ \ \ \ \ \ \ \ \ \ \ \ \ \ \ \ \ + \left[ \tilde{\sigma}_{j-1,j}(t) \tilde{\rho} \tilde{\sigma}_{k-1,k}(t') - \tilde{\rho} \tilde{\sigma}_{k-1,k}(t') \tilde{\sigma}_{j-1,j}(t) \right] \langle \tilde{\Gamma}_k^\dagger(t') \tilde{\Gamma}_j^\dagger(t) \rangle_R \nonumber \\
	&& \ \ \ \ \ \ \ \ \ \ \ \ \ \ \ \ \ + \left[ \tilde{\sigma}_{k,k-1}(t') \tilde{\rho} \tilde{\sigma}_{j,j-1}(t) - \tilde{\sigma}_{j,j-1}(t) \tilde{\sigma}_{k,k-1}(t') \tilde{\rho} \right] \langle \tilde{\Gamma}_j(t) \tilde{\Gamma}_k(t') \rangle_R \nonumber \\
	&& \ \ \ \ \ \ \ \ \ \ \ \ \ \ \ \ \ + \left[ \tilde{\sigma}_{j,j-1}(t) \tilde{\rho} \tilde{\sigma}_{k,k-1}(t') - \tilde{\rho} \tilde{\sigma}_{k,k-1}(t') \tilde{\sigma}_{j,j-1}(t) \right] \langle \tilde{\Gamma}_k(t') \tilde{\Gamma}_j(t) \rangle_R \nonumber \\
	&& \ \ \ \ \ \ \ \ \ \ \ \ \ \ \ \ \ + \left[ \tilde{\sigma}_{k,k-1}(t') \tilde{\rho} \tilde{\sigma}_{j-1,j}(t) - \tilde{\sigma}_{j-1,j}(t) \tilde{\sigma}_{k,k-1}(t') \tilde{\rho} \right] \langle \tilde{\Gamma}_j^\dagger(t) \tilde{\Gamma}_k(t') \rangle_R \nonumber \\
	&& \ \ \ \ \ \ \ \ \ \ \ \ \ \ \ \ \ + \left[ \tilde{\sigma}_{j-1,j}(t) \tilde{\rho} \tilde{\sigma}_{k,k-1}(t') - \tilde{\rho} \tilde{\sigma}_{k,k-1}(t') \tilde{\sigma}_{j-1,j}(t) \right] \langle \tilde{\Gamma}_k(t') \tilde{\Gamma}_j^\dagger(t) \rangle_R \nonumber \\
	&& \ \ \ \ \ \ \ \ \ \ \ \ \ \ \ \ \ + \left[ \tilde{\sigma}_{k-1,k}(t') \tilde{\rho} \tilde{\sigma}_{j,j-1}(t) - \tilde{\sigma}_{j,j-1}(t) \tilde{\sigma}_{k-1,k}(t') \tilde{\rho} \right] \langle \tilde{\Gamma}_j(t) \tilde{\Gamma}_k^\dagger(t') \rangle_R \nonumber \\
	&& \ \ \ \ \ \ \ \ \ \ \ \ \ \ \ \ \ + \left. \left[ \tilde{\sigma}_{j,j-1}(t) \tilde{\rho} \tilde{\sigma}_{k-1,k}(t') - \tilde{\rho} \tilde{\sigma}_{k-1,k}(t') \tilde{\sigma}_{j,j-1}(t) \right] \langle \tilde{\Gamma}_k^\dagger(t') \tilde{\Gamma}_j(t) \rangle_R \right\} , \label{A14}
\end{eqnarray}
\end{widetext}
where
\begin{eqnarray}
 \langle \tilde{\Gamma}_j^\dagger(t)\tilde{\Gamma}_k^\dagger(t') \rangle_R &=& {\mathrm{Tr}}_R \left[ R(0) \tilde{\Gamma}_j^\dagger(t)\tilde{\Gamma}_k^\dagger(t') \right] , \nonumber \\
\langle \tilde{\Gamma}_k^\dagger(t')\tilde{\Gamma}_j^\dagger(t) \rangle_R &=& {\mathrm{Tr}}_R \left[ R(0) \tilde{\Gamma}_k^\dagger(t')\tilde{\Gamma}_j^\dagger(t) \right] , \nonumber \\
\langle \tilde{\Gamma}_j(t)\tilde{\Gamma}_k(t') \rangle_R &=& {\mathrm{Tr}}_R \left[ R(0) \tilde{\Gamma}_j(t)\tilde{\Gamma}_k(t') \right] , \nonumber\\
\langle \tilde{\Gamma}_k(t')\tilde{\Gamma}_j(t) \rangle_R &=& {\mathrm{Tr}}_R \left[ R(0) \tilde{\Gamma}_k(t')\tilde{\Gamma}_j(t) \right] , \nonumber \\
\langle \tilde{\Gamma}_j^\dagger(t)\tilde{\Gamma}_k(t') \rangle_R &=& {\mathrm{Tr}}_R \left[ R(0) \tilde{\Gamma}_j^\dagger(t)\tilde{\Gamma}_k(t') \right] ,\nonumber\\
\langle \tilde{\Gamma}_k(t')\tilde{\Gamma}_j^\dagger(t) \rangle_R &=& {\mathrm{Tr}}_R \left[ R(0) \tilde{\Gamma}_k(t')\tilde{\Gamma}_j^\dagger(t) \right] , \nonumber \\
\langle \tilde{\Gamma}_j(t)\tilde{\Gamma}_k^\dagger(t') \rangle_R &=& {\mathrm{Tr}}_R \left[ R(0) \tilde{\Gamma}_j(t)\tilde{\Gamma}_k^\dagger(t') \right] , \nonumber \\
\langle \tilde{\Gamma}_k^\dagger(t')\tilde{\Gamma}_j(t) \rangle_R &=& {\mathrm{Tr}}_R \left[ R(0) \tilde{\Gamma}_k^\dagger(t')\tilde{\Gamma}_j(t) \right]  \nonumber 
\end{eqnarray}
are {\em two-time correlation functions}. These two-time correlation functions are expected to vanish for $t - t'$ larger than a time scale $\tau_c \propto \hbar/k_B T$ called the {\em correlation time} of the reservoir. Thus, the integral in Eq. (\ref{A14}) is nonzero only for times $0 \leq t - t' \leq \tau_c$, and for $t'$ outside this time interval, $\tilde{\rho}(t')$ should not affect $\dot{\tilde{\rho}}(t)$ at time $t$. As discussed in Ref. [\onlinecite{Carmichael}] we assume that the time scale for significant change of the system is much larger than the reservoir correlation time; then the Markov approximation Eq. (\ref{A9}) holds.

Since we also suppose that the reservoir is in thermal equilibrium, the correlation functions with terms $\langle a_\alpha^\dagger a_\alpha^\dagger \rangle_R$ or $\langle a_\alpha a_\alpha \rangle_R$ vanish, and the terms $\langle a_\alpha^\dagger a_\alpha \rangle_R$ and $\langle a_\alpha a_\alpha^\dagger \rangle_R$ give
\begin{eqnarray}
	{\mathrm{Tr}}_R \left[ R(0) a_\alpha^\dagger a_\alpha \right] &=& \bar{n}(\omega_\alpha,T) ,  \label{A16-1}\\
{\mathrm{Tr}}_R \left[ R(0) a_\alpha a_\alpha^\dagger \right] &=& 1 + \bar{n}(\omega_\alpha,T) . \label{A16}
\end{eqnarray}
Here $\bar{n}(\omega_\alpha,T)$ has the meaning of mean photon number for an oscillator with frequency $\omega_\alpha$ in thermal equilibrium at temperature $T$,
\begin{equation}
\bar{n}(\omega_\alpha,T)=\frac{e^{-\hbar\omega_\alpha / k_B T}}{1 - e^{-\hbar\omega_\alpha / k_B T}}.
\end{equation}

We change the summation over the reservoir levels in the nonvanishing correlation functions to integrations by introducing a density of states $G(\omega)$, and use $\tau \equiv t - t'$ as a new variable in Eq. (\ref{A14}). By substituting the nonvanishing correlation functions and their corresponding system operators in equation (\ref{A14}), the Markovian master equation in the interaction picture is explicitly written as follows (define the detuning $\Delta \equiv \omega_{10} - \omega_{21}$)
\begin{widetext}
\begin{eqnarray}
	\dot{\tilde{\rho}} &=& (\sigma_{10}\tilde{\rho}\sigma_{01} - \sigma_{01}\sigma_{10}\tilde{\rho}) \int_0^t d\tau \int_0^\infty d\omega e^{i(\omega - \omega_{10})\tau} G(\omega)|f_1(\omega)|^2 \bar{n}(\omega,T) \nonumber \\
	&+& (\sigma_{01}\tilde{\rho}\sigma_{10} - \tilde{\rho}\sigma_{10}\sigma_{01}) \int_0^t d\tau \int_0^\infty d\omega e^{i(\omega - \omega_{10})\tau} G(\omega)|f_1(\omega)|^2 [\bar{n}(\omega,T)+1] \nonumber \\
	&+& (\sigma_{21}\tilde{\rho}\sigma_{01} - \sigma_{01}\sigma_{21}\tilde{\rho}) \int_0^t d\tau \int_0^\infty d\omega e^{i(\omega - \omega_{21})\tau} e^{-i\Delta t} G(\omega)f_1(\omega) f_2^\ast(\omega) \bar{n}(\omega,T) \nonumber \\
	&+& (\sigma_{01}\tilde{\rho}\sigma_{21} - \tilde{\rho}\sigma_{21}\sigma_{01}) \int_0^t d\tau \int_0^\infty d\omega e^{i(\omega - \omega_{21})\tau} e^{-i\Delta t} G(\omega) f_1(\omega) f_2^\ast(\omega) [\bar{n}(\omega,T)+1] \nonumber \\
	&+& (\sigma_{01}\tilde{\rho}\sigma_{10} - \sigma_{10}\sigma_{01}\tilde{\rho}) \int_0^t d\tau \int_0^\infty d\omega e^{-i(\omega - \omega_{10})\tau} G(\omega)|f_1(\omega)|^2 [\bar{n}(\omega,T)+1] \nonumber \\
	&+& (\sigma_{10}\tilde{\rho}\sigma_{01} - \tilde{\rho}\sigma_{01}\sigma_{10}) \int_0^t d\tau \int_0^\infty d\omega e^{-i(\omega - \omega_{10})\tau} G(\omega)|f_1(\omega)|^2 \bar{n}(\omega,T) \nonumber \\
	&+& (\sigma_{12}\tilde{\rho}\sigma_{10} - \sigma_{10}\sigma_{12}\tilde{\rho}) \int_0^t d\tau \int_0^\infty d\omega e^{-i(\omega - \omega_{21})\tau} e^{i\Delta t} G(\omega) f_1^\ast(\omega) f_2(\omega) [\bar{n}(\omega,T)+1] \nonumber \\
	&+& (\sigma_{10}\tilde{\rho}\sigma_{12} - \tilde{\rho}\sigma_{12}\sigma_{10}) \int_0^t d\tau \int_0^\infty d\omega e^{-i(\omega - \omega_{21})\tau} e^{i\Delta t} G(\omega)f_1^\ast(\omega) f_2(\omega) \bar{n}(\omega,T) \nonumber \\
	&+& (\sigma_{10}\tilde{\rho}\sigma_{12} - \sigma_{12}\sigma_{10}\tilde{\rho}) \int_0^t d\tau \int_0^\infty d\omega e^{i(\omega - \omega_{10})\tau} e^{i\Delta t} G(\omega)f_1^\ast(\omega) f_2(\omega) \bar{n}(\omega,T) \nonumber \\
	&+& (\sigma_{12}\tilde{\rho}\sigma_{10} - \tilde{\rho}\sigma_{10}\sigma_{12}) \int_0^t d\tau \int_0^\infty d\omega e^{i(\omega - \omega_{10})\tau} e^{i\Delta t} G(\omega) f_1^\ast(\omega) f_2(\omega) [\bar{n}(\omega,T)+1] \nonumber \\
	&+& (\sigma_{21}\tilde{\rho}\sigma_{12} - \sigma_{12}\sigma_{21}\tilde{\rho}) \int_0^t d\tau \int_0^\infty d\omega e^{i(\omega - \omega_{21})\tau} G(\omega)|f_2(\omega)|^2 \bar{n}(\omega,T) \nonumber \\
	&+& (\sigma_{12}\tilde{\rho}\sigma_{21} - \tilde{\rho}\sigma_{21}\sigma_{12}) \int_0^t d\tau \int_0^\infty d\omega e^{i(\omega - \omega_{21})\tau} G(\omega)|f_2(\omega)|^2 [\bar{n}(\omega,T)+1] \nonumber \\
	&+& (\sigma_{01}\tilde{\rho}\sigma_{21} - \sigma_{21}\sigma_{01}\tilde{\rho}) \int_0^t d\tau \int_0^\infty d\omega e^{-i(\omega - \omega_{10})\tau} e^{-i\Delta t} G(\omega) f_1(\omega) f_2^\ast(\omega) [\bar{n}(\omega,T)+1] \nonumber \\
	&+& (\sigma_{21}\tilde{\rho}\sigma_{01} - \tilde{\rho}\sigma_{01}\sigma_{21}) \int_0^t d\tau \int_0^\infty d\omega e^{-i(\omega - \omega_{10})\tau} e^{-i\Delta t} G(\omega)f_1(\omega) f_2^\ast(\omega) \bar{n}(\omega,T) \nonumber \\
	&+& (\sigma_{12}\tilde{\rho}\sigma_{21} - \sigma_{21}\sigma_{12}\tilde{\rho}) \int_0^t d\tau \int_0^\infty d\omega e^{-i(\omega - \omega_{21})\tau} G(\omega)|f_2(\omega)|^2 [\bar{n}(\omega,T)+1] \nonumber \\
	&+& (\sigma_{21}\tilde{\rho}\sigma_{12} - \tilde{\rho}\sigma_{12}\sigma_{21}) \int_0^t d\tau \int_0^\infty d\omega e^{-i(\omega - \omega_{21})\tau} G(\omega)|f_2(\omega)|^2 \bar{n}(\omega,T) . \label{A17}
\end{eqnarray}
\end{widetext}
Since $t$ is the typical time scale for changes in $\tilde{\rho}$, which is much longer than reservoir correlation time $\tau_c$, we can approximately extend the upper limit of the time integral to infinity, and then we evaluate the integrals in Eq. (\ref{A17}) as follows (without loss of generality, assume $f_1$ and $f_2$ are real)
\begin{widetext}
\begin{eqnarray}
	&& \int_0^t d\tau \int_0^\infty d\omega e^{\pm i(\omega - \omega_{10})\tau} G(\omega)f_1^2(\omega) \bar{n}(\omega,T) = \pi G(\omega_{10}) f_1^2(\omega_{10})\bar{n}(\omega_{10},T) , \nonumber \\
	&& \int_0^t d\tau \int_0^\infty d\omega e^{\pm i(\omega - \omega_{21})\tau} G(\omega)f_2^2(\omega) \bar{n}(\omega,T) = \pi G(\omega_{21}) f_2^2(\omega_{21})\bar{n}(\omega_{21},T) , \nonumber \\
	&& \int_0^t d\tau \int_0^\infty d\omega e^{\pm i(\omega - \omega_{10})\tau} G(\omega)f_1(\omega) f_2(\omega) \bar{n}(\omega,T) = \pi G(\omega_{10}) f_1(\omega_{10}) f_2(\omega_{10}) \bar{n}(\omega_{10},T) , \nonumber \\
	&& \int_0^t d\tau \int_0^\infty d\omega e^{\pm i(\omega - \omega_{21})\tau} G(\omega)f_1(\omega) f_2(\omega) \bar{n}(\omega,T) = \pi G(\omega_{21}) f_1(\omega_{21}) f_2(\omega_{21}) \bar{n}(\omega_{21},T) . \label{A19}
\end{eqnarray}
\end{widetext}
Since $\omega_{10}\approx\omega_{21}$ ($\omega_{10} = 2\pi\times 8.135$ GHz and $\omega_{21} = 2\pi\times 7.975$ GHz), we can make the approximations $G(\omega_{10})\approx G(\omega_{21}) \equiv G$, $f_{1(2)}(\omega_{10})\approx f_{1(2)}(\omega_{21}) \equiv f_{1(2)}$, and $\bar{n}(\omega_{10},T) \approx \bar{n}(\omega_{21},T) \equiv \bar{n}$.

Combining equation (\ref{A17}) with equation (\ref{A19}), and using $\sigma_{01}\sigma_{21} = \sigma_{21}\sigma_{01} = \sigma_{10}\sigma_{12} = \sigma_{12}\sigma_{10} = 0$, we get the master equation in following form
\begin{eqnarray}
	\dot{\tilde{\rho}} &=& \frac{\Gamma_{10}}{2}\bar{n}(2 \sigma_{10} \tilde{\rho} \sigma_{01} - \sigma_{00}\tilde{\rho} - \tilde{\rho}\sigma_{00})  \nonumber \\
&+&
\frac{\Gamma_{10}}{2}(\bar{n} + 1) (2 \sigma_{01}\tilde{\rho}\sigma_{10} - \sigma_{11}\tilde{\rho} - \tilde{\rho}\sigma_{11}) \nonumber \\
	&+& \frac{\Gamma_{21}}{2} \bar{n} (2 \sigma_{21}\tilde{\rho}\sigma_{12} - \sigma_{11}\tilde{\rho} - \tilde{\rho}\sigma_{11}) \\
&+&
\frac{\Gamma_{21}}{2}(\bar{n} + 1) (2 \sigma_{12}\tilde{\rho}\sigma_{21} - \sigma_{22}\tilde{\rho} - \tilde{\rho}\sigma_{22}) \nonumber \\
	&+& \kappa e^{-i\Delta t} \sigma_{01}\tilde{\rho}\sigma_{21} + \kappa \bar{n} e^{-i\Delta t} (\sigma_{01}\tilde{\rho}\sigma_{21} + \sigma_{21}\tilde{\rho}\sigma_{01}) \nonumber \\
	&+& \kappa e^{i\Delta t} \sigma_{12}\tilde{\rho}\sigma_{10} + \kappa \bar{n} e^{-i\Delta t} (\sigma_{12}\tilde{\rho}\sigma_{10} + \sigma_{10}\tilde{\rho}\sigma_{12}) ,\nonumber
\label{A20a}
\end{eqnarray}
where the energy relaxation rates $\Gamma_{10}\equiv 2\pi G f_1^2$, $\Gamma_{21}\equiv 2\pi G f_2^2$, and $\kappa \equiv \sqrt{\Gamma_{10}\Gamma_{21}}$.

Transforming back to the Schr\"{o}dinger picture, and taking the zero temperature limit $\bar{n}\rightarrow 0$, we finally obtain the Markovian master equation
\begin{equation}
	\dot{\rho} = - \ i[H_S, \rho] \ + \ e^{-iH_St}\dot{\tilde{\rho}}e^{iH_St} \nonumber
\end{equation}
in the final form [see also Eq. (\ref{eq_relax})]
\begin{eqnarray}
\dot{\rho}	&=& -i[H_S, \rho] + \kappa (\sigma_{01}\rho \sigma_{21} + \sigma_{12}\rho \sigma_{10}) \nonumber \\
	&& + \frac{\Gamma_{10}}{2} (2 \sigma_{01}\rho \sigma_{10} - \sigma_{11}\rho - \rho \sigma_{11}) \nonumber \\
	&& + \frac{\Gamma_{21}}{2} (2 \sigma_{12}\rho \sigma_{21} - \sigma_{22}\rho - \rho \sigma_{22}) .
\label{A20}
\end{eqnarray}
In a doubly-rotating frame, terms with $\kappa$ average out to zero on the time scale of our measurements, due to the factor $e^{\pm i\delta t}$ (see Sec. \ref{master_eq}).


\section{Pure dephasing}
\label{appendixb}

In this Appendix we derive the dephasing part of the Liouvillean for a three-level system, as
 introduced in Section \ref{master_eq}. We also present numerical results bearing on the validity of the approximation $\gamma_{21} \approx  \gamma_{20}+\gamma_{10}$. For our system we suggest that an appropriate model for dephasing could be that of a harmonic oscillator with fluctuating frequencies in addition to uncorrelated
noise due to (energy-conserving) virtual transitions with the environment.

For two-level quantum systems, it is known that dephasing is
caused by longitudinal noises ({\it i.e.} along $\sigma_z$), which induce fluctuations in the transition
frequency.
Although for three-level systems we do not have a Bloch sphere picture to help our intuition, we can still
develop the theory by using the matrices $\sigma_{jk}$ defined in Appendix \ref{appendixa}. 

We consider the ladder three-level Hamiltonian Eq. (\ref{A2}), and
\begin{equation}
H_{fl}(t) = \hbar \delta\omega_{10}(t)|1\rangle\langle 1| + \hbar [\delta\omega_{21} (t) +\delta\omega_{10} (t)]
|2\rangle\langle 2|
\end{equation}
describing random fluctuations of transition frequencies.
The total Hamiltonian in matrix form reads
\begin{eqnarray}
& & H_{tot}(t) = H_S + H_{fl}(t) \nonumber \\
& &= \hbar \left[ \begin{array}{ccc}
	0 & 0 & 0 \\
	0 & \omega_{10} + \delta\omega_{10}(t) & 0 \\
	0 & 0 & \omega_{21} + \omega_{10} +  \delta\omega_{20} (t)
\end{array} \right] ,
\end{eqnarray}
where we define $\delta\omega_{20}(t) =  \delta\omega_{21}(t) + \delta\omega_{10}(t)$.

In the interaction picture, the dynamics of the three-level system's density operator,
\begin{equation}
\tilde\rho(t) = \exp(iH_S t) \rho(t) \exp(-iH_S t),
\end{equation}
follows from the Schr\"odinger equation and a perturbation expansion to the second order in noises,
\begin{eqnarray}
i\hbar\dot{\tilde\rho} &=& \left[ \tilde{H}_{fl}(t), \tilde\rho(t) \right] \nonumber \\
&\approx& \left[ \tilde{H}_{fl}(t), \tilde\rho(0)+\frac{1}{i\hbar}\int_0^t dt' \left[ \tilde{H}_{fl}(t'), \tilde\rho(t') \right] \right] \nonumber \\
&\approx& \left[ \tilde{H}_{fl}(t), \tilde\rho(0) \right] \nonumber \\
&&+ \frac{1}{i\hbar} \int_0^t dt' \left[ \tilde{H}_{fl}(t), \left[ \tilde{H}_{fl}(t'), \tilde\rho(t') \right] \right] , \label{eq_B2}
\end{eqnarray}
where $\tilde{H}_{fl}(t) = \exp(iH_S t) H_{fl}(t) \exp(-iH_S t)$. The double commutator $\left[ \tilde{H}_{fl}(t), \left[ \tilde{H}_{fl}(t'), \tilde\rho(t')
\right] \right]$ from the last line of Eq. (\ref{eq_B2}) can be calculated and has the following matrix form
\begin{widetext}
\begin{equation}
 \left[ \begin{array}{ccc}
0 & \tilde\rho_{01}(t') [\delta\omega_{10}(t) \delta\omega_{10}(t')] &
\tilde\rho_{02}(t') [\delta\omega_{20}(t) \delta\omega_{20}(t')] \\
\tilde\rho_{10}(t') [\delta\omega_{10}(t) \delta\omega_{10}(t')] & 0 &
\tilde\rho_{12}(t') [\delta\omega_{10}(t) - \delta\omega_{20}(t)]
[\delta\omega_{10}(t') - \delta\omega_{20}(t')] \\
\tilde\rho_{20}(t') [\delta\omega_{20}(t) \delta\omega_{20}(t')] &
\tilde\rho_{21}(t') [\delta\omega_{10}(t) - \delta\omega_{20}(t)]
[\delta\omega_{10}(t') - \delta\omega_{20}(t')] & 0
\end{array} \right], \nonumber
\end{equation}\nonumber
\end{widetext}
Then, after averaging over the fluctuations, Eq. (\ref{eq_B2}) can be rewritten in the following matrix form,
\begin{widetext}
\begin{eqnarray}
\dot{\tilde\rho}(t) &\approx& -\frac{i}{\hbar}\left[ \tilde{H}_{fl}(t), \tilde\rho(0) \right] \nonumber \\
&& + \frac{1}{2} \left[ \begin{array}{ccc}
	0 & -\gamma_{10}^\varphi \tilde\rho_{01}(t) & -\gamma_{20}^\varphi \tilde\rho_{02}(t) \\
	-\gamma_{10}^\varphi \tilde\rho_{10}(t) & 0 & -(\gamma_{10}^\varphi + \gamma_{20}^\varphi - S_{12} - S_{21}) \tilde\rho_{12}(t) \\
	-\gamma_{20}^\varphi \tilde\rho_{20}(t) & -(\gamma_{10}^\varphi + \gamma_{20}^\varphi - S_{12} - S_{21}) \tilde\rho_{21}(t) & 0
\end{array} \right] , \label{eq_B3}
\end{eqnarray}
\end{widetext}
where we have as usual the density matrix element as $\tilde\rho_{jk}(t) = \langle j| \tilde\rho(t) |k\rangle$ and we made  the approximation $\tilde\rho(t') \approx \tilde\rho(t)$ (see also Appendix \ref{appendixa})).  The dephasing rates are defined as
\begin{eqnarray}
\gamma_{10}^\varphi &=& \frac{2}{\hbar^2} \int_0^t dt' \langle \delta\omega_{10}(t)\delta\omega_{10}(t')\rangle ,\label{eq_BB}\\
\gamma_{20}^\varphi &=& \frac{2}{\hbar^2} \int_0^t dt' \langle \delta\omega_{20}(t)\delta\omega_{20}(t')\rangle ,\label{eq_BBB}
\end{eqnarray}
and the cross spectral densities as
\begin{equation}
S_{12} = \frac{2}{\hbar^2}\int_0^t dt' \langle \delta\omega_{10}(t) \delta\omega_{20}(t') \rangle = S_{21} .\label{eq_BBBB}
\end{equation}
Let us now define define a parameter $\varepsilon = S_{12} + S_{21}$, which parametrizes the effect of correlations between levels 1 and 2. We can now write the dephasing terms in Eq. (\ref{eq_B3}) as
\begin{eqnarray}
{\cal L}_{\rm dep}[\tilde\rho] &=& \sum_{j=1,2}\frac{\gamma_{j0}^\varphi}{2} \left( 2\sigma_{jj}\tilde\rho\sigma_{jj} - \sigma_{jj}\tilde\rho - \tilde\rho\sigma_{jj} \right) \nonumber \\
&& \ \ + \frac{\varepsilon}{2} \left( \sigma_{11}\tilde\rho\sigma_{22} + \sigma_{22}\tilde\rho\sigma_{11} \right) ,
\end{eqnarray}
or equivalently in  the form used in the main part of the paper Eq. (\ref{eq_dephs}),
\begin{equation}
{\cal L}_{\rm dep} [\tilde{\rho}] =
- \sum_{j,k \in \{0,1,2\};j\neq k} \frac{\gamma^{\varphi}_{jk}}{2} \sigma_{jj}\tilde{\rho}\sigma_{kk},
\end{equation}
where $\gamma^{\varphi}_{20}=\gamma^{\varphi}_{21}+\gamma^{\varphi}_{10}-\varepsilon$.

\begin{figure}[h]
\includegraphics[width=8.5cm]{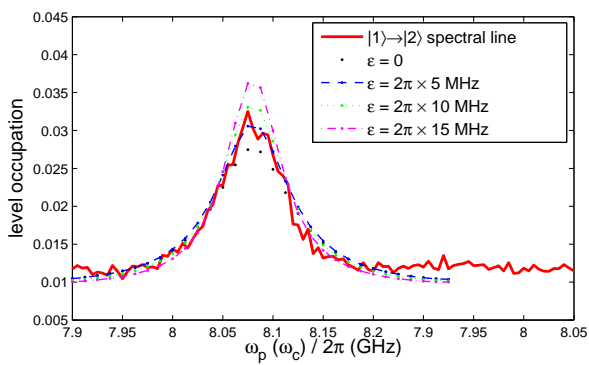}
\caption{Numerical fitting of $|1\rangle\rightarrow|2\rangle$ spectral line (the same as traces shown in Fig. \ref{fig_spectro}) with different vales of $\varepsilon$. The following parameters (as in Table \ref{table}) are used for this simulation: $\Gamma_{10} = 2\pi\times 7$MHz, $\Gamma_{21} = 2\pi\times 11$MHz, $\gamma_{10}^\varphi = 2\pi\times 7$MHz and $\gamma_{20}^\varphi = 2\pi\times 16$MHz.}
\label{fig_spectro_epsilon}
\end{figure}


To summarize,
the master equation including both energy relaxation and dephasing is (now back in the Schr\"odinger picture)
\begin{eqnarray}
\dot{\rho} &=& -i [H_S, \rho] + \kappa \left( \sigma_{01}\rho\sigma_{21} + \sigma_{12}\rho\sigma_{10} \right) \nonumber \\
&& + \frac{\Gamma_{10}}{2} \left( 2 \sigma_{01}\rho\sigma_{10} - \sigma_{11}\rho - \rho\sigma_{11} \right) \nonumber \\
&& + \frac{\Gamma_{21}}{2} \left( 2 \sigma_{12}\rho\sigma_{21} - \sigma_{22}\rho - \rho\sigma_{22} \right) \nonumber \\
&& + \frac{\gamma_{10}^\varphi}{2} \left( 2 \sigma_{11}\rho\sigma_{11} - \sigma_{11}\rho - \rho\sigma_{11} \right) \nonumber \\
&& + \frac{\gamma_{20}^\varphi}{2} \left( 2 \sigma_{22}\rho\sigma_{22} - \sigma_{22}\rho - \rho\sigma_{22} \right) \nonumber \\
&& + \frac{\varepsilon}{2} \left( \sigma_{11}\rho\sigma_{22} + \sigma_{22}\rho\sigma_{11} \right) ,
\end{eqnarray}
which we use to fit the experimental data with fixed dephasing rates, relaxation rates and different values of $\varepsilon$.

As shown in Fig. \ref{fig_spectro_epsilon}, a finite value of $\varepsilon$ of about $2\pi\times 5$MHz provides the best fitting (the (blue) dashed dotted curves). Values such as $\varepsilon = 0$ MHz and $\varepsilon = 2\pi\times 10$MHz provide reasonably good fittings, while larger values, such as $\varepsilon = 2\pi\times 15$ MHz result in much worse fittings. We have also simulated the Autler-Townes spectra of Fig. \ref{fig_Autler} and we see no significant difference for values $0\leq \varepsilon \leq 2\pi\times 10$MHz, while for larger values the fittings become worse. Changing the value of other parameters (such as $\gamma_{20}^{\varphi}$) does not produce better results either. We conclude that in our experiment $\varepsilon$ is approximately $2\pi\times 5$MHz.

Next, we suggest that the dephasing in this system can be understood as originating from two processes. Since the system is an oscillator with only weak anharmonicity, the first process is the fluctuation of the frequency $\omega_{0}$ of the oscillator , which depends on external fluctuating parameters such as the bias flux $\Phi_{\rm dc}$ (as defined in Section \ref{multi_level}). The second source of dephasing are virtual transitions between the qubit and the environment (electromagnetic degrees of freedom, two-level systems, {\it etc.}). The first process can be experimentally controlled in a more straigthforward way, for example by filtering of the bias lines and by using magnetic-flux pinning substrates, while the second process is of a more fundamental nature. For both processes, we can use the general formalism presented above to find the expressions for the
dephasing Liouvilleans.

The first process will be characterized by a single dephasing parameter $\gamma_{\rm HO}^{\varphi}$, since we note that
\begin{eqnarray}
\delta{\omega}_{10}(t) &=& \frac{\partial \omega_{10}}{\partial \Phi_{\rm dc}}\delta\Phi_{\rm dc}(t)
                 \approx  \frac{\partial \omega_{0}}{\partial \Phi_{\rm dc}}\delta\Phi_{\rm dc}(t)                 \approx \delta{\omega}_{21}(t),\nonumber \\
\delta{\omega}_{20}(t) &=& \frac{\partial (\omega_{10}+ \omega_{21})}{\partial \Phi_{\rm dc}}\delta\Phi_{\rm dc}(t)
\approx  2\delta{\omega}_{10}(t) \approx 2\delta{\omega}_{21}(t),\nonumber
\end{eqnarray}
where the derivatives are taken at the flux bias point and $\delta\Phi_{\rm dc}(t)$ is the fluctuation in the dc-bias magnetic field (see also Section \ref{multi_level}). From Eqs. (\ref{eq_B3}, \ref{eq_BB}, \ref{eq_BBB}, and \ref{eq_BBBB}) we have $\gamma_{\rm HO}^{\varphi} = 2 \hbar^{-2} \int_0^t dt' \langle \delta\omega_{0}(t)\delta\omega_{0}(t')\rangle$ and
\begin{equation}
{\cal L}_{\rm HO}[\tilde\rho] = \frac{1}{2} \left[ \begin{array}{ccc}
	0 & -\gamma_{\rm HO}^{\varphi}\tilde\rho_{01} & -4\gamma_{\rm HO}^{\varphi}\tilde\rho_{02} \\
	-\gamma_{\rm HO}^{\varphi}\tilde\rho_{10} & 0 & -\gamma_{\rm HO}^{\varphi}\tilde\rho_{12} \\
	-4\gamma_{\rm HO}^{\varphi}\tilde\rho_{20} & -\gamma_{\rm HO}^{\varphi}\tilde\rho_{21} & 0
\end{array} \right] . \label{eq_B9}
\end{equation}
Alternatively, we can start with the known expression for the dephasing of the harmonic oscillator \cite{Milburn}
\begin{equation}
{\cal L}_{\rm HO}[\tilde\rho] = \frac{\gamma_{\rm HO}^{\varphi}}{2}\left[ 2a^\dag a \tilde\rho a^\dag a - (a^\dag a)^2 \tilde\rho - \tilde\rho (a^\dag a)^2 \right] , \label{eq_B6}
\end{equation}
and truncate the creation and annihilation operators of the harmonic oscillator to the lowest three levels,
\begin{equation}
a^\dag = \left[ \begin{array}{ccc}
	0 & 0 & 0 \\
	1 & 0 & 0 \\
	0 & \sqrt{2} & 0
\end{array} \right] ,
\ \ \ \ a = \left[ \begin{array}{ccc}
	0 & 1 & 0 \\
	0 & 0 & \sqrt{2} \\
	0 & 0 & 0
\end{array} \right] .
\end{equation}
In this three-level Hilbert space, Eq. (\ref{eq_B6}) is identical to Eq. (\ref{eq_B9}).

The second process can be characterized by two parameters, $\Gamma^{\varphi}_{1}$ and $\Gamma^{\varphi}_{2}$, and, for processes involving virtual transitions with the environment
the Liouvillean can be obtained again from the general expressions Eqs. (\ref{eq_B3}, \ref{eq_BB}, \ref{eq_BBB}, and \ref{eq_BBBB}) for the particular case of zero interlevel correlations. The resulting expression is well-known also from quantum optics \cite{EIT,Dephasing},
\begin{equation}
{\cal L}_{\rm virt}[\tilde{\rho}] = \sum_{j=1,2}\frac{\Gamma_j^\varphi}{2}\left( 2\sigma_{jj}\tilde{\rho}\sigma_{jj} - \sigma_{jj}\tilde{\rho} - \tilde{\rho}\sigma_{jj} \right).
\end{equation}
With these assumptions in mind, we then have ${\cal L}_{\rm deph}[\tilde{\rho}] = {\cal L}_{\rm HO}[\tilde{\rho}]
+ {\cal L}_{\rm virt}[\tilde{\rho}]$, which results in $\gamma^{\varphi}_{10} = \gamma_{\rm HO}^{\varphi} + \Gamma_{1}^\varphi$,
$\gamma^{\varphi}_{20} = 4 \gamma_{\rm HO}^{\varphi} + \Gamma^\varphi_{2}$, and
$\gamma^{\varphi}_{21} = \gamma_{\rm HO}^{\varphi} + \Gamma_{1}^\varphi+ \Gamma^\varphi_{2}$.
This means that $\varepsilon = 4\gamma_{\rm HO}^{\varphi}$, and if we take
$\varepsilon = 2\pi \times 5$ MHz we find $\gamma_{\rm HO}^{\varphi}= 2\pi \times 1.25$ MHz,
$\Gamma_{1}^{\varphi} = 2\pi \times 5.75$ MHz, and $\Gamma_{2}^{\varphi} = 2\pi \times 11$ MHz.


\section{The steady-state density matrix}
\label{appendixc}

To find the elements of the steady-state density matrix, it is possible to solve analytically (using Mathematica) the equation $\dot{\rho}^{\rm (st)} = 0$ for the case $\Delta_{c}=0$. The complete expressions are complicated, but they can be simplified and put in a form amenable to physical interpretation in certain limits. In this appendix we consider the approximation $\Omega_{p}^{2}\ll \gamma_{20}^2, \gamma_{10}^2, \gamma_{10}\gamma_{20}\ll\Omega_{c}^{2}$. We find:
\begin{widetext}
\begin{eqnarray}
\rho_{10}^{(\rm st)}&=&\rho_{01}^{(\rm st)*}=\frac{(2 \Delta_{p} - i \gamma_{20} )\Omega_{p}}{-4 \Delta_{p}^2+2 i \Delta_{p} (\gamma_{10}+\gamma_{20})+\Omega_{c}^2}, \label{B110}\\
\rho_{20}^{(\rm st)}&=&\rho_{02}^{(\rm st)*}=-\frac{\Omega_{c} \Omega_{p}}{-4 \Delta_{p}^2+2 i \Delta_{p} (\gamma_{10}+\gamma_{20})+\Omega_{c}^2} , \label{B19}\\
\rho_{21}^{(\rm st)}&=&\rho_{12}^{(\rm st)*}=\frac{i \left(4 \Delta_{p}^2 \Gamma_{21} (\Gamma_{10}-\gamma_{10})+2 i \Delta_{p} \Gamma_{10} \Omega_{c}^2-\Gamma_{21} (\Gamma_{10}+\gamma_{20}) \Omega_{c}^2\right) \Omega_{p}^2}{\Gamma_{10} \Omega_{c} \left(-4 \Delta_{p}^2+\Omega_{c}^2\right)^2},
\label{B18} \\
\rho_{11}^{(\rm st)}&=&\frac{\left(4 \Delta_{p}^2 \gamma_{10}+\gamma_{20} \Omega_{c}^2\right) \Omega_{p}^2}{\Gamma_{10} \left(-4 \Delta_{p}^2+\Omega_{c}^2\right)^2},\label{B111}\\
\rho_{22}^{(\rm st)}&=&-\frac{\left(4 \Delta_{p}^2 (\Gamma_{10}-\gamma_{10})-(\Gamma_{10}+\gamma_{20}) \Omega_{c}^2\right) \Omega_{p}^2}{\Gamma_{10} \left(-4 \Delta_{p}^2+\Omega_{c}^2\right)^2},\label{B112}\\
\rho_{00}^{(\rm st)}&=&1-\rho_{11}^{(\rm st)}-\rho_{22}^{(\rm st)}.\label{B113}
\end{eqnarray}
\end{widetext}
As a quick consistency check, one sees that the elements $\rho_{11}^{(\rm st)}$ Eq. (\ref{B111}) and $\rho_{01}^{(\rm st)}$ Eq. (\ref{B110}) can be obtained from the expressions Eq. (\ref{eq_occupation_prob}) and Eq. (\ref{rho10}) if we restrict ourselves to $\gamma_{20}^2, \gamma_{10}^2, \gamma_{10}\gamma_{20}\ll\Omega_{c}^{2}$. Using the expressions above, we can calculate the fidelity Eq. (\ref{fid}) with respect to the dark state in the rotating frame,
\begin{eqnarray}
{\cal F}_{|D\rangle} [\rho^{(\rm st)}] &=& \frac{\cos 2\Theta}{2}(\rho_{00}^{(\rm st)}-\rho_{22}^{(\rm st)}) - \frac{\sin 2\Theta}{2}(\rho_{20}^{(\rm st)}+\rho_{02}^{(\rm st)}) \nonumber \\
&& + \frac{1}{2}(1-\rho_{11}^{(\rm st)}).
\end{eqnarray}
Inserting now the expressions of Eqs. (\ref{B110}-\ref{B113}) and using the approximations $\Omega_{p}^{2}\ll \gamma_{20}^2, \gamma_{10}^2, \gamma_{10}\gamma_{20}\ll\Omega_{c}^{2}$, it can be shown by algebraic calculations in Mathematica, that the fidelity is close to 1 for any detuning $\Delta_{p}$.

More insight into the structure of the state can be obtained, however, in the resonant case, $\Delta_{p}=0$. In this situation the density matrix elements Eqs. (\ref{B110}-\ref{B113}) become
\begin{eqnarray}
\rho_{10}^{(\rm st)}&=&\rho_{01}^{(\rm st)*}=\frac{i \gamma_{20} \Omega_{p}}{\Omega_{c}^2} , \label{B10} \\
\rho_{20}^{(\rm st)}&=&\rho_{02}^{(\rm st)*}=-\frac{\Omega_{p}}{\Omega_{c}} \\
\rho_{21}^{(\rm st)}&=&\rho_{12}^{(\rm st)*}=\frac{-i\Gamma_{21} (\Gamma_{10}+\gamma_{20})\Omega_{p}^2}{\Gamma_{10}\Omega_{c}^3} \\
\rho_{11}^{(\rm st)}&=&\frac{\gamma_{20}  \Omega_{p}^2}{\Gamma_{10}\Omega_{c}^2}, \label{B11} \\
\rho_{22}^{(\rm st)}&=&\frac{(\Gamma_{10}+\gamma_{20})  \Omega_{p}^2}{\Gamma_{10} \Omega_{c}^2}, \\
\rho_{00}^{(\rm st)}&=&1-\rho_{11}^{(\rm st)}-\rho_{22}^{(\rm st)}.
\end{eqnarray}
Now for an ideal dark state $|D\rangle$ in the limit $\Omega_{p}\ll \Omega_{c}$ we have  $|D\rangle = (1-\Omega_{p}^2/2\Omega_{c}^{2})|0\rangle - (\Omega_{p}/\Omega_{c})|2\rangle$ and the corresponding matrix elements are
\begin{eqnarray}
\rho_{10}^{(|D\rangle)}&=&\rho_{01}^{*(|D\rangle)}=0 ,\label{B10}\\
\rho_{20}^{(|D\rangle)}&=&\rho_{02}^{*(|D\rangle)}=-\frac{\Omega_{p}}{\Omega_{c}},\\
\rho_{21}^{(|D\rangle)}&=&\rho_{12}^{*(|D\rangle)}=0 ,\\
\rho_{11}^{(|D\rangle)}&=&0,\\
\rho_{22}^{(|D\rangle)}&=&\frac{\Omega_{p}^2}{\Omega_{c}^2}, \\
\rho_{00}^{(|D\rangle)}&=&1-\rho_{22}^{(|D\rangle)}.
\end{eqnarray}

A simple visual comparison between the two sets of elements shows that they coincide up to first order in $\Omega_{p}/\Omega_{c}$ (assuming $\gamma_{20}/\Omega_{c}$ is of the same order or smaller). Even a better coincidence, up to second order in $\Omega_{p}/\Omega_{c}$, could in principle be reached provided that $(\gamma_{20}/\Omega_{c})\ll(\Omega_{p}/\Omega_{c})^2$, $(\gamma_{20}/\Gamma_{10})\leq (\Omega_{p}/\Omega_{c})$, and $(\Gamma_{21}/\Omega_{c})\sim (\Gamma_{10}/\Omega_{c})\sim (\Omega_{p}/\Omega_{c})$. This agrees with the intuition that the effect introduces deviations from the ideal dark state second order in $\Omega_{p}/\Omega_{c}$ are the decoherence effects  associated with state $|2\rangle$ (dephasing and relaxation to state $|1\rangle$). In conclusion, it is not surprising that states which are reasonable close to true dark states are produced in this experiment.


\end{document}